\documentclass[aps,floatfix,groupaddress]{revtex4}
\usepackage{graphicx,amssymb,amsfonts,amsmath,color,bm,hyperref,subfig}
\usepackage{xcolor}
\usepackage{hyphenat}
\hypersetup{
    colorlinks,
    linkcolor={blue},
    citecolor={blue},
    urlcolor={black}
}

\begin{document}
\title{Threshold response and bistability in gene regulation by small noncoding RNA}
\author{Sutapa Mukherji}
\affiliation{Department of Protein Chemistry and Technology, 
Central Food Technological Research Institute, 
Mysore-570 020, India}

\date{\today}
\begin{abstract}

In this paper, we study through mathematical modelling the combined effect of    transcriptional and translational regulation 
by proteins and  small noncoding RNAs (sRNA)  in a genetic feedback motif  that 
 has  an important role in the  survival  of   {\it E.coli} under stress associated with  oxygen and energy 
 availability.  We show that  subtle changes in this motif can bring in drastically different effects  on the gene expression.  In particular, 
 we show that  a threshold response in  the gene expression changes to a bistable response as  the regulation on  sRNA synthesis  
 or degradation is altered.   These results are obtained under  deterministic conditions. Next, we  study how the gene expression is 
 altered by additive and multiplicative noise which might arise  due to probabilistic occurrences of  different biochemical events. 
Using the Fokker-Planck formulation,  we  obtain   steady state probability distributions for  sRNA  concentration  for the network 
motifs displaying bistability. The probability distributions  are  found to be  bimodal with two peaks at low and high concentrations 
of sRNAs. We further study the variations in the  probability distributions  under different values of noise strength and  correlations. 
The results presented here might be of interest for  designing synthetic network for artificial control.

 \end{abstract}

\maketitle
\section{Introduction}
\label{sec:introduction}
Translational and transcriptional regulations are two important regulation 
strategies of gene expression. In transcriptional regulation, a protein 
regulator, binds to the DNA and activates or represses the mRNA synthesis \cite{alon}. 
The translational regulation is often accomplished by small noncoding regulatory 
 RNA (sRNA) molecules which primarily function through sequence specific 
base-pairing with the target mRNAs \cite{gottesmanrev}. 
Experimental studies show that sRNAs can modulate the ribosome 
binding to the target mRNA by either sequestering the ribosome binding site 
causing translational repression \cite{masse} or by exposing the ribosome binding 
site  facilitating  translation \cite{majdalani}. In addition, there are 
increasing number of studies which demonstrate that under 
different contexts,  sRNA, through 
base pairing with target mRNA, can prevent or facilitate RNase E mediated 
degradation of mRNA affecting mRNA stability \cite{papenfort}.  Like protein regulators, 
sRNAs regulate 
multiple target genes and their own synthesis is regulated by other 
transcription factors. Since sRNAs are not translated, it is believed that 
it is beneficial for the cell to have such regulatory RNAs. Further due to 
the small structure, it is possible that  the synthesis of such molecules 
is more energetically efficient than the synthesis of the protein regulators.

Gene expression  networks consist of certain types of
frequently occurring subnetworks, called  
network motifs \cite{mangan-alon,rosenfeld, shen-orr,bose}. 
 Frequent occurrences of these  network motifs in gene 
expression networks indicate  that these motifs are important due to the
specific advantages they  provide to the cell \cite{alon}. 
 Most of the earlier research has 
focussed on  network motifs involving protein mediated transcriptional regulation.
Recent studies on regulatory 
interactions with sRNA,  reveal that a large number of  network  motifs  function  
using dual strategies  consisting  of both transcriptional and translational 
regulation.  The protein-mediated transcriptional regulation is fundamentally 
different from the sRNA mediated post-transcriptional regulation. 
For example, while in protein mediated transcriptional repression 
 the gene expression is completely shutdown by the  binding of the transcriptional repressor 
 to the DNA,  the effectiveness of  the post-transcriptional repression by sRNA depends largely on the sRNA  and mRNA 
 synthesis rates and their binding affinity. 
Recent studies \cite{hwa,hwa2} reveal that sRNA mediated regulation 
can give rise to several interesting features such as  threshold-linear response in gene expression, attenuation 
of noise in protein synthesis etc.   which are, in general,
 unexpected from protein mediated transcriptional regulation.  In view of these results, 
 it appears  important to understand  how different minimal designs of 
network motifs involving sRNA and protein regulators   can bring in drastically different consequences.

One of the important consequences  of various nonlinear interactions in network motifs 
 is bistability which implies the  existence of two possible steady state solutions 
for same parameter sets.  These two solutions are stable  solutions, corresponding to, 
say, high or low expression of a gene, and these solutions 
appear along  with  an  unstable solution of an intermediate value.
Thus, over a range of a suitable parameter value, for example, a specific interaction strength, 
the signal-response curve (in this case,  the bifurcation diagram) consists of  two stable branches 
of solution separated by an intermediate unstable branch. Due to the presence of an 
intermediate  unstable solution, a continuous change  from a low to a high expression 
state is not possible.  The cell is thus locked into either  a high or a low expression regime 
in an irreversible manner.  So far, a large number of examples of bistable switches 
 have  been found  in different contexts. Some of these are 
 responsible for controlling alternative life style of phage $\lambda$ \cite{arkin,kobiler}, 
 cell cycle progression \cite{novak,sha}, cell fate determination in sea urchin \cite{davidson} 
  etc. It is believed that complex interactions among various components at the  transcriptional or 
  post-transcriptional level and  conservation laws 
are responsible for such ultra sensitivity \cite{igoshin}. 
Another kind of ultra sensitivity  
is found in sRNA-mediated post-translational regulation. 
 In \cite{hwa,hwa2},  it was shown that for a 
  target transcription rate below a threshold value set by the sRNA transcription rate, the gene 
expression is completely silenced while  beyond the  threshold value, 
there is a smooth transition to a different regime where the gene expression  
increases linearly with the difference between the  mRNA and sRNA transcription rates.  
The nonlinear interactions  through which sRNAs  regulate the target mRNA concentrations seem to be responsible 
for such  threshold response  in the steady-state. 

\begin{figure}
\centering
\begin{minipage}{0.45\textwidth}
\centering
 \includegraphics[height=1.4\textwidth]{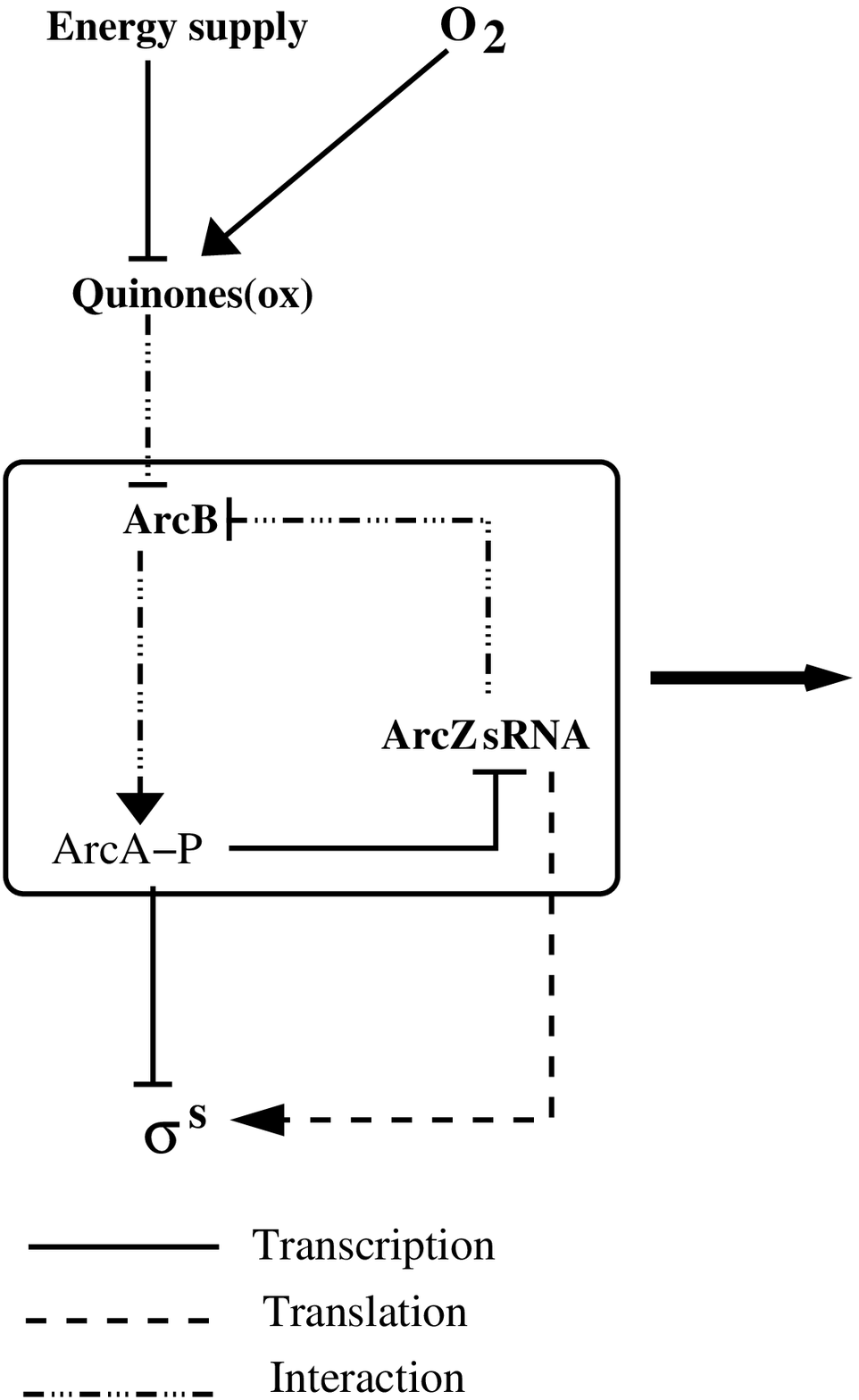} \\ {\bf  (A)}
 \end{minipage}
 \centering
\begin{minipage}{0.45\textwidth}
\centering
\includegraphics[height=1.2\textwidth]{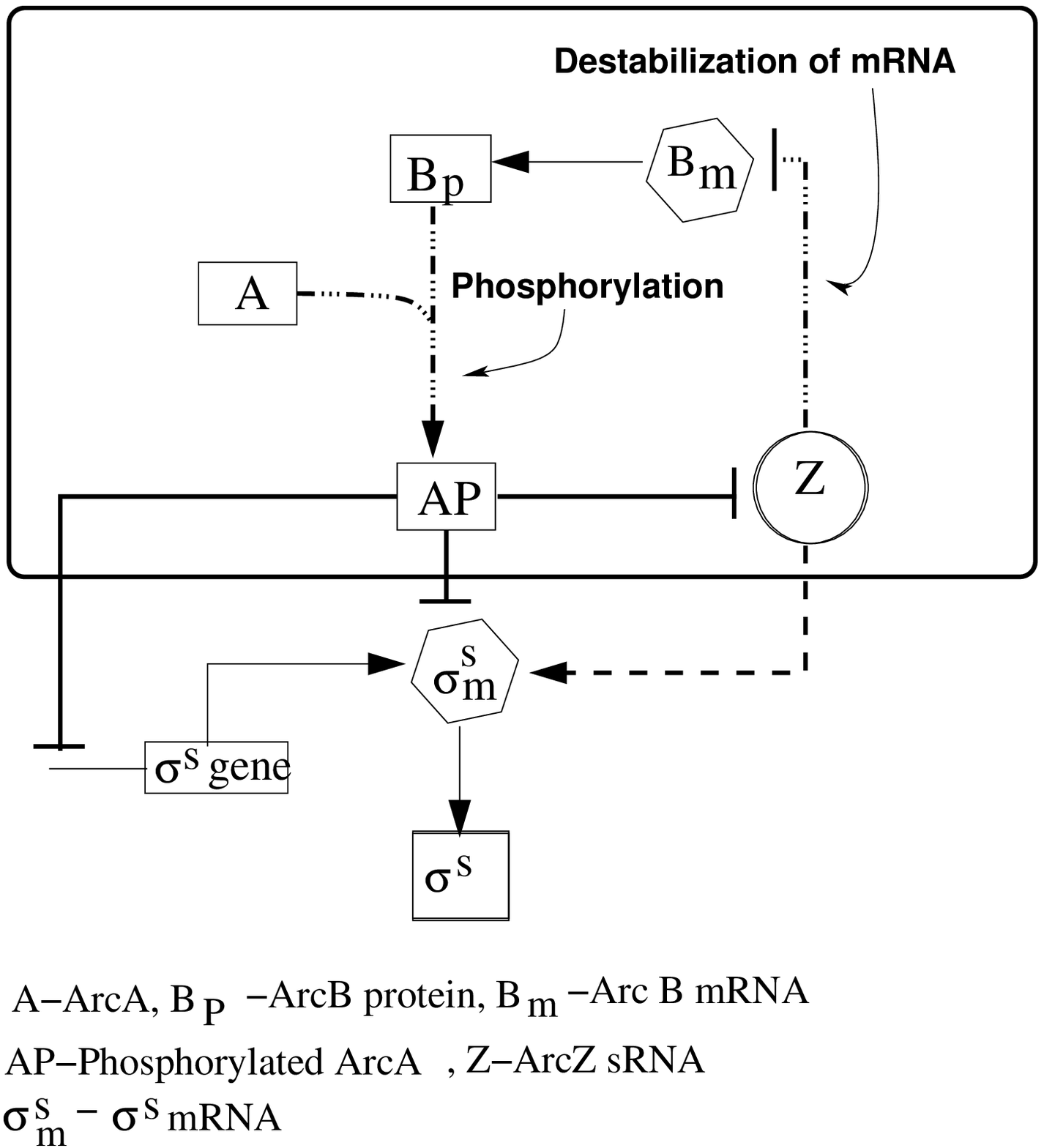} {\bf (B)}
 \end{minipage}
 \caption{The network responsible 
 for $\sigma^s$ regulation based on oxygen and energy availability in the environment. 
 Lines ending with a bar represent repression. Arrowed lines represent activation. 
   (A) Oxidized quinone inhibits autophosphorylation 
 of the sensor kinase ArcB. ArcB upon autophosphorylation, phosphorylates ArcA. Phosphorylated ArcA is a transcriptional inhibitor 
that  represses the transcription  of ArcZ sRNA and $\sigma^s$ mRNA.  ArcZ sRNA destabilises  ArcB mRNA.  The part of the network inside the 
 big rectangle is of our present interest.  (B) Some of the intermediate steps of (A) are shown in detail. 
 For this figure as well as for the next figure, rectangles, hexagons   and circles represent protein regulators, 
mRNA and sRNA molecules, respectively. Model I is based on this network. The phosphorylated A ( AP), in a monomeric form, 
represses transcription of Z-sRNA. In model II (see figure (\ref{fig:2scenarios}A)),  
the transcription of Z-sRNA is inhibited by a homodimer of AP.}
\label{fig:samplefig1}
\end{figure}

Here we are interested in  exploring how small changes in the network architecture influence 
 the steady-state properties of the network. In particular, we are interested in 
network motifs that involve dual strategies i.e. two different regulation mechanisms involving protein and sRNA regulators. 
The specific motif with which we begin our analysis 
 is a subnetwork of  a larger regulatory network that is 
responsible for the response of the bacteria under stress due to oxygen and energy 
availability \cite{hengge,gottesman,mika}.  As we show below, this subnetwork   under certain circumstances shows 
threshold response in the steady-state.  Beginning with this subnetwork, we introduce 
small modifications in the network architecture to study  how 
the threshold response behaviour is affected  due to  minor  alteration in the network architecture. 
 In this network, the  regulation  is activated through the 
environment sensing by the histidine sensor kinase  ArcB (see figure \ref{fig:samplefig1}A).
Under high  oxygen and  
energy starvation conditions, quinones are oxidised. These oxidised quinones 
inhibit the autophosphorylation activity of ArcB which, upon autophosphorylation, 
phosphorylates ArcA. The phosphorylated ArcA represses the synthesis of $\sigma^s$ in two ways. 
It represses  the transcription of $\sigma^s$ mRNA directly and also influences the 
translation of $\sigma^s$ by repressing the transcription of  ArcZ sRNA which 
 activates  $\sigma^s$ translation. 
Further, in a feed-back mechanism, ArcZ sRNA destabilizes ArcB mRNA. 
The destabilisation of ArcB affects the  ArcA phosphorylation and this, in turn, promotes the 
 synthesis of ArcZ sRNA. 
 Overall, the subnetwork represents a complex regulation mechanism  with 
  dual strategies involving protein and sRNA mediated regulation.  
In the low oxygen and high energy condition,  activated ArcA represses ArcZ synthesis  thereby leading to high   
ArcB concentration  and  hence 
further activation of ArcA.  Conversely, in case of high oxygen   and  low energy, there is  reduced activation of ArcA 
leading to 
enhanced expression of ArcZ which leads to down-regulation of ArcB level. 
Thus the  switching from one mode to the other is governed by ArcA phosphorylation which 
is driven by environment sensing of ArcB sensor kinase.  
In this work,  we  primarily  focus  on the regulation of the 
ArcZ sRNA level  and how the  details of the network architecture may  impact
ArcZ concentration levels.   The ArcZ concentration level is crucial  for $\sigma^s$ translation. However, in order to understand 
 $\sigma^s$ regulation, it is necessary to consider 
  another  layer of regulation of $\sigma^s$ at the proteolysis level by the proteolytic factor RssB 
which itself is regulated based on the environmental conditions. Thus  
it would be of interest to find  how  the changes in sRNA level  would finally impact the  
$\sigma^s$ level.
For sake of generality and also for future convenience with the mathematical equations, we 
 refer the reader to table (\ref{table-notation})  displaying  the short notations  for various components. 
 \begin{table}[ht]
\caption{Notations used in the text}
\centering
\begin{tabular} {|c| c| c|}
\hline\hline
Protein/mRNA/sRNA & Short Form & Concentration \\
\hline
ArcZ sRNA & Z-sRNA & [Z]  \\
\hline
ArcB mRNA& B-mRNA &$[B_m]$   \\
\hline
ArcB  Protein & B-Protein & $[B_p]$ \\
\hline
ArcA protein & A-protein   & $[A]$ \\
\hline
Complex of ArcB and ArcA & Complex of A- and B-protein  & $[AB_p]$ \\
\hline
Phosphorylated ArcA & Phosphorylated A & $[AP]$ \\
\hline
Complex of ArcB mRNA and ArcZ sRNA & Complex of B-mRNA and Z-sRNA & $[B_m Z]$\\
\hline
$\sigma^s$ mRNA & $\sigma^s$ mRNA & $[\sigma^sm]$ \\
\hline
Complex of $\sigma^s$ mRNA and  ArcZ sRNA & Complex of $\sigma^s$-mRNA and Z-sRNA &  $[\sigma^smZ]$ \\
\hline
\end{tabular}\label{table-notation}\\
\end{table}

\begin{figure}[ht!]
  \centering
   \includegraphics[height=0.6\textwidth]{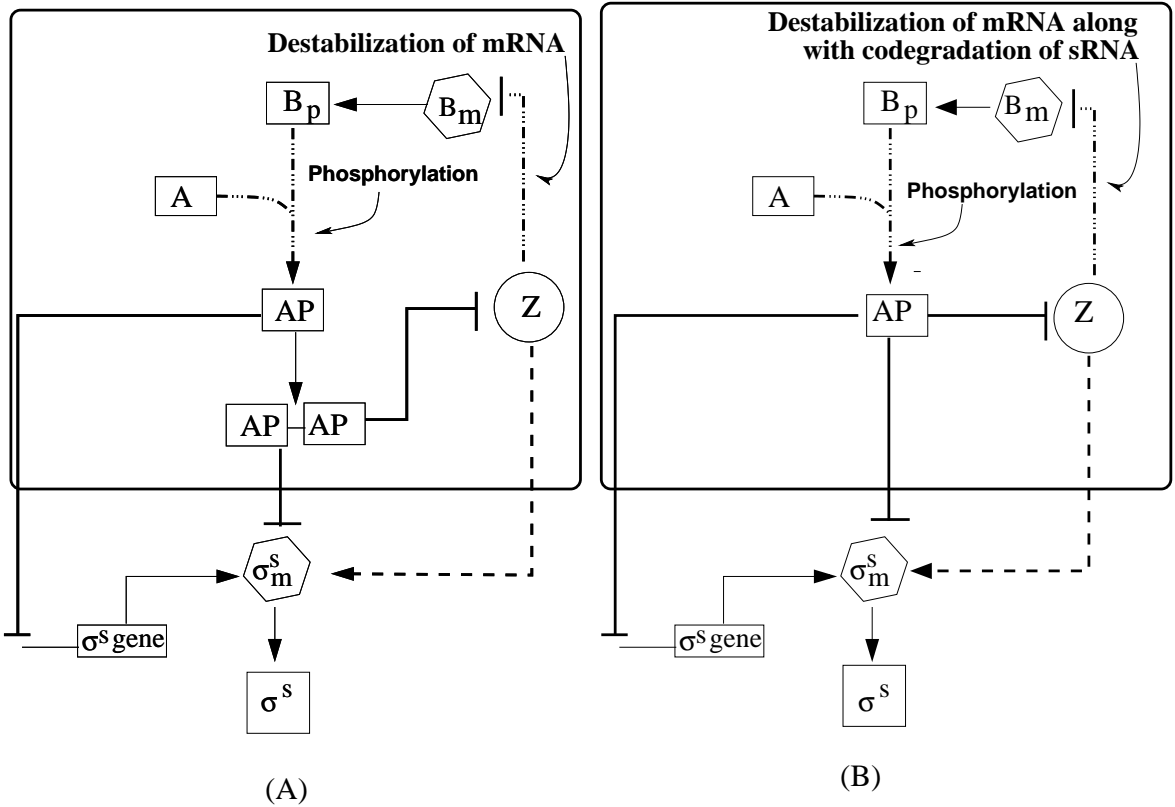}  
\caption{(A) The network that describes model II. A  homodimeric form of phosphorylated A (AP) inhibits the transcription of 
Z-sRNA (Z). The rest of the regulation mechanism is same as that of figure (\ref{fig:samplefig1}B). (B) The network that represents model III. 
This network is same as model I (figure (\ref{fig:samplefig1}B)) except 
that here Z-sRNA (Z) codegardes with B-mRNA ($B_m$). No codegradation of sRNA is considered in model II (in panel (A)) of this figure. }
\label{fig:2scenarios}
\end{figure}

Being motivated by this  basic architecture  of the subnetwork, we consider  three  possible 
 regulation scenarios. 
In model I  (shown in  figure (\ref{fig:samplefig1}B)),  
the phosphorylated protein regulator (phosphorylated ArcA or  AP),  binds to the  DNA to repress the synthesis 
of  Z-sRNA (Z).   In this case, the Z-sRNA concentration shows a threshold response  as 
the Z-sRNA mediated B-mRNA  ($B_m$) degradation rate is increased. 
In model II ( shown in  figure (\ref{fig:2scenarios}A)), the phosphorylated protein 
regulator (phosphorylated ArcA or AP)  forms   a homodimer for  
repressing  Z-sRNA synthesis.  In  case of ArcA, 
 it is known that ArcA belongs to the OmpR/ PhoB subfamily of response regulators. Earlier 
it has been hypothesized   that the members of this subfamily 
use a common mechanism of dimerization for regulation. 
Further, additional structural studies on ArcA reveal that  oligomerization of  ArcA, in general,  
might be important for the ArcA mediated regulation \cite{stock}. Model II proposed here might 
account for such possibilities in a motif. The threshold behaviour seen in case of model I 
disappears completely in case of model II with Z-sRNA concentration showing a bistable response as 
the Z-sRNA mediated B-mRNA degradation is increased. 
In  case of model I and model II,  we assume that the Z-sRNA  molecule 
upon destabilising B-mRNA, returns back to the  system for further activity. The third model (referred as model III) (see figure (\ref{fig:2scenarios}B)) 
is similar to  model I  apart from a subtle  difference that  sRNAs   co-degrade along with the   mRNAs.   
The degradation of mRNA upon formation of the sRNA-mRNA complex can be of  different types including one-to-one, partial or no 
codegradation of sRNA  along with the target mRNA \cite{hwa2}. Thus, model I (as well as model II) and model III represent the cases of no 
codegradation and one-to-one 
codegradation of sRNA, respectively.   We find that   like model II, model III also shows bistable response in the sRNA concentration as 
sRNA-mRNA interaction is changed.
 While  cooperativity seems to be responsible for bistability in model II,  it is the 
 one-to-one codegradation of sRNA    that causes 
 bistability in model III. Further, we find that  for model III, the bistable response is much more robust compared to model II with a   wider bistable  region.  
 These results are obtained upon analysing  deterministic (noise-free) equations describing the time evolution of  concentrations  of various components. 

The gene expression is inherently noisy and this leads to population heterogeneity in a  bacterial colony. In case of bistability, there are
 possibilities of noise mediated  switching  between the low and high expression states \cite{bosecomk,veening}.  For example, a switching from a low to a high expression 
 state happens if  due to the noise, the protein concentration has a value higher than the  threshold set by the concentration of the unstable solution. 
  Since the two stable gene expression states (low and high expression states) are connected through noise driven,
 stochastic transitions, the high or low expression state of a cell must be described probabilistically. 
  The  second part of our analysis is aimed at finding the probability  distributions of sRNA concentrations for model II and model III  
 for which a  deterministic analysis predicts the possibility of bistability in the steady-state.

 \section{Models}
 In this section, we present a detailed description of the models in terms of the rate equations describing the time evolutions 
 of various components. 
 
 The rate equations describing the  time evolutions of B-mRNA and  the B-protein  concentrations are
\begin{eqnarray}
\frac{d[B_m]}{dt}=-\gamma_{bm} [B_m]+r_{bm}-k_{bz}^+ [B_m][Z]  \ \ \  {\rm and} \label{diffeqnbm}\\
\frac{d[B_p]}{dt}=-k_c^+ [B_p][A]+k_c^-[AB_p]+k_p^+[AB_p]-\gamma_{bp} [B_p]+r_{bp} [B_m]. \label{diffeqnbp}
\end{eqnarray}
Here, $k_{bz}^+$ denotes the rate of Z-sRNA  mediated destabilization of B-mRNA;
$r_{bm}$ and $\gamma_{bm}$ denote the rate of synthesis and degradation,
respectively, of B-mRNA; 
 $r_{bp}$ and $\gamma_{bp}$ represent the rate of synthesis and degradation, respectively,  
  of  B-protein.  First, second and third  terms of equation (\ref{diffeqnbp}) represent complex formation 
  between $A$ and $B$ protein molecules at rate $k_c^+$, 
  dissociation of the complex ($[AB_p]$)at rate $k_c^-$, and release of B protein  
  from the complex $[AB_p]$  upon phosphorylation 
  of $A$  at rate $k_p^+$ (see Appendix for further details). 
  The steady-state concentration levels of B-mRNA and B-protein  can be obtained by equating
  the above rate equations to zero. This leads to 
  \begin{eqnarray}
&&[B_m]=\frac{r_{bm}}{k_{bz}^+[Z]+\gamma_{bm}} \ \ {\rm and} \label{steadybm}\\
&& [B_p]=\frac{r_{bp}}{\gamma_{bp}}[B_m].\label{steadybp}
\end{eqnarray}

 For the entire dynamics, above equations must be combined with the  rate equations for
the concentrations of Z-sRNA and  concentrations of two types of  mRNA-sRNA complexes, $[B_mZ]$ and $[\sigma^smZ]$.    
The rate equations for Z-sRNA  depends on the specific model we consider. In case of model I, the time evolution of the 
sRNA concentration is described by 
\begin{eqnarray}
\frac{d[Z]}{dt}=\beta_z \frac{G_z^{\rm tot}}{1+k_z [AP]}-k_{bz}^+ [B_m][Z]+k_{bz}^-[B_mZ]-k_{mz}^+
[\sigma^sm][Z]+
k_{mz}^- [\sigma^smZ]-\gamma_z [Z],\label{diffeqnz0}
\end{eqnarray}
where $\beta_z$ represents the synthesis rate of Z-sRNA by the active gene. The first term in 
the above expression indicates  transcriptional inhibition by the protein regulator, phosphorylated A (AP). In case of model I, this 
transcriptional inhibition happens through the monomeric  form of the regulator, AP.  The second  term represents  the 
complex formation between Z-sRNA and B-mRNA at a rate $k_{bz}^+$.  Upon the formation of this complex, B-mRNA is degraded at rate $k_{bz}^-$. 
Since in model I,   Z-sRNA does not undergo codegradation with B-mRNA,  free Z-sRNA is recycled back into the system 
as it degrades    B-mRNA.  The fourth and the fifth term represent  complex formation and dissociation of $\sigma^s$-mRNA and Z-sRNA at rates 
$k_{mz}^+$ and $k_{mz}^-$, respectively.  The last term represents the degradation of Z-sRNA. 
Assuming  the synthesis rate of $\sigma^s$-mRNA from the active state of the gene to be $\beta_\sigma$, we have 
\begin{eqnarray}
\frac{d[\sigma^sm]}{dt}=\beta_\sigma \frac{G_\sigma^{\rm tot}}{1+k_\sigma [AP]}+k_{mz}^- [\sigma^smZ]-
k_{mz}^+[\sigma^sm][Z]-\gamma_m [\sigma^sm]\label{diffeqnsigmam}\\
\frac{d[B_mZ]}{dt}=k_{bz}^+ [B_m] [Z]-k_{bz}^-[B_mZ]\label{diffeqnbz}\\
\frac{d[\sigma^smZ]}{dt}=k_{mz}^+ [\sigma^sm][Z]-k_{mz}^-[\sigma^sm Z], \label{diffeqnsigmamz}
\end{eqnarray}
where $k_\sigma= \frac{k_\sigma^+}{k_\sigma^-}$ and 
 $k_z=\frac{k_z^+}{k_z^-}$ (see  appendix \ref{synthesis}). The first term in equation (\ref{diffeqnsigmam}) again represents
  transcriptional inhibition in the synthesis of $\sigma^s$-mRNA by the monomeric 
 form of the inhibitor, AP.  The last term represents the degradation of $\sigma^s$ mRNA at rate $\gamma_m$. 
 The remaining terms in equations (\ref{diffeqnsigmam}), (\ref{diffeqnbz}) and (\ref{diffeqnsigmamz})  represent formation and dissociation 
 of different types of complexes. 

In model II, we  assume that the phosphorylated A protein (AP) binds the DNA in the homodimer form. 
The time evolution equations  for various concentrations  except for the equation  for
Z-sRNA (i.e. equation (\ref{diffeqnz0}))  are the same as those of Model I.
 The time evolution equation  for Z-sRNA  for model II  is 
\begin{eqnarray}
\frac{d[Z]}{dt}=\beta_z \frac{G_z^{\rm tot}}{1+k_z [AP]^2}-k_{bz}^+ [B_m][Z]+k_{bz}^-[B_mZ]-k_{mz}^+
[\sigma^sm][Z]+
k_{mz}^- [\sigma^smZ]-\gamma_z [Z].\label{diffeqnz1}
\end{eqnarray} 
Here the first term indicating transcriptional repression of Z-sRNA by the $[AP]$ homodimer is different from that 
of Model I.  

In case of Model III, we consider co-degradation of Z-sRNA as it destabilises B-mRNA. 
The time evolution equations in this case are similar to  those of Model I except that  for this case 
the time evolution of Z-sRNA is modified as 
\begin{eqnarray}
\frac{d[Z]}{dt}=\beta_z \frac{G_z^{\rm tot}}{1+k_z [AP]}-k_{bz}^+ [B_m][Z]-k_{mz}^+
[\sigma^sm][Z]+k_{mz}^- [\sigma^smZ]-\gamma_z [Z].\label{diffeqnz2}
\end{eqnarray}
Further, since Z-sRNA  co-degrades with B-mRNA,  equation (\ref{diffeqnbz})
is no longer relevant for this model.

\section{Deterministic analysis}
\label{sec:deter}
\subsection{Model I}


Using the equilibrium solution for $[AP]$ (see appendix \ref{const-A}) and equations 
(\ref{steadybm}) and (\ref{steadybp})  for  $[B_p]$ and $[B_m]$, we have  from (\ref{diffeqnz0})
\begin{eqnarray}
\gamma_z [Z]\left[1+\left( \frac{r_{bp}}{\gamma_{bp}}\right)\frac{k_{AP} k_z r_{bm} [A]}{k_{bz}^+[Z]+\gamma_{bm}}\right]=\beta_z G_z^{\rm tot}. \label{zsolution}
\end{eqnarray}
Being a quadratic equation in $[Z]$, equation (\ref{zsolution})  has two solutions of which the physically acceptable solution is  
\begin{eqnarray}
[Z]=[Z^*]=\frac{1}{2 k^+_{bz}}\bigg\{-(\gamma_{bm}+K-\frac{\beta_z}{\gamma_z} G_z^{\rm tot} k_{bz}^+)+ \big [(\gamma_{bm}+K-
\frac{\beta_z}{\gamma_z} G_z^{\rm tot} k_{bz}^+)^2+\nonumber\\
4 \frac{\gamma_{bm}}{\gamma_z}\beta_z G_z^{\rm tot} k_{bz}^+\big ]^{1/2}\bigg\},
\label{steadyz}
\end{eqnarray}
where $K=\frac{k_z k_{AP} r_{bm} r_{bp}[A]}{\gamma_{bp}}$.  Here, $k_{AP}=\frac{k_p k_c}{1+k_p^+/k_c^-}$ with $k_p=\frac{k_p^+}{k_p^-}$, $k_c=\frac{k_c^+}{k_c^-}$. 
Although the  presence of bistability is ruled out in this case, as we discuss below,  the solution indicates  a threshold response for the Z-sRNA concentration.

\subsubsection{Numerical Results}
Equations (\ref{diffeqnbm}), (\ref{diffeqnbp}), (\ref{diffeqnz0}), (\ref{diffeqnsigmam}), (\ref{diffeqnbz}) and  (\ref{diffeqnsigmamz}) 
can be solved numerically  to see how  the concentrations $[Z]$ and $[B_p]$ evolve with time.   A  quasi steady-state approximation
is made here by considering  steady-state for the phosphorylation kinetics part. 
Various parameter values used for these solutions are listed in table (\ref{table-one}). These are the average 
 values  obtained from references \cite{alon, shimoni}. sRNA synthesis 
rate is considered to be $5$ times larger than the  mRNA synthesis rates.  Values of $k_z^+$ and $k_\sigma^+$ are the 
average values obtained from \cite{alon}. 
Since the detachment rates typically have wide variations depending  on the bond strength, we 
assume $k_z^-,\ k_\sigma^-=1.5\ {\rm sec}^{-1}$ ($>1\ {\rm sec}^{-1}$) \cite{alon}.  Using the average  values for 
phosphorylation and dephosphorylation rates (see reference \cite{miller}), we find $k_{\rm AP}=0.004\ {\rm molecule}^{-1}$. 
For low and high  phosphorylation 
rates, we choose $k_{\rm AP}=0.001,\  0.1\  {\rm molecule}^{-1}$, respectively.  Further, all the results are obtained with a
 fixed concentration of A-protein (see Appendix \ref{const-A}). 

\begin{figure}[!htb]
  \minipage{0.5\textwidth}
 \includegraphics[height=0.55\textwidth]{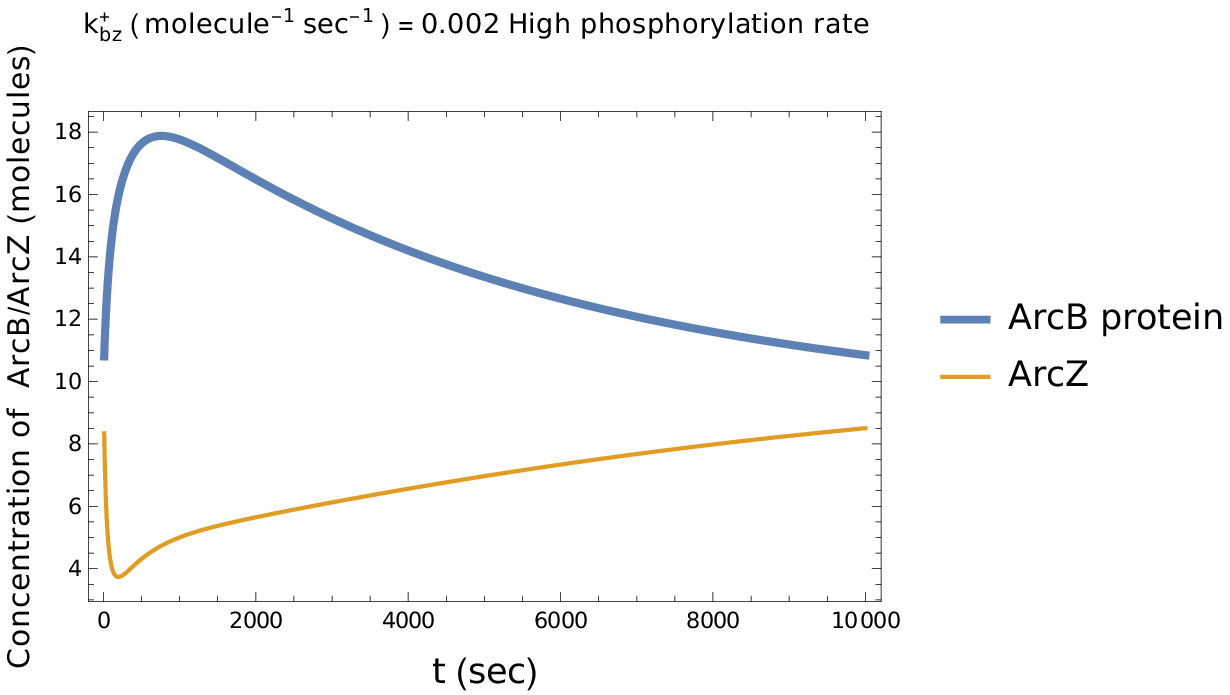}
 \endminipage
 \minipage{0.5\textwidth}
 \includegraphics[height=0.55\textwidth]{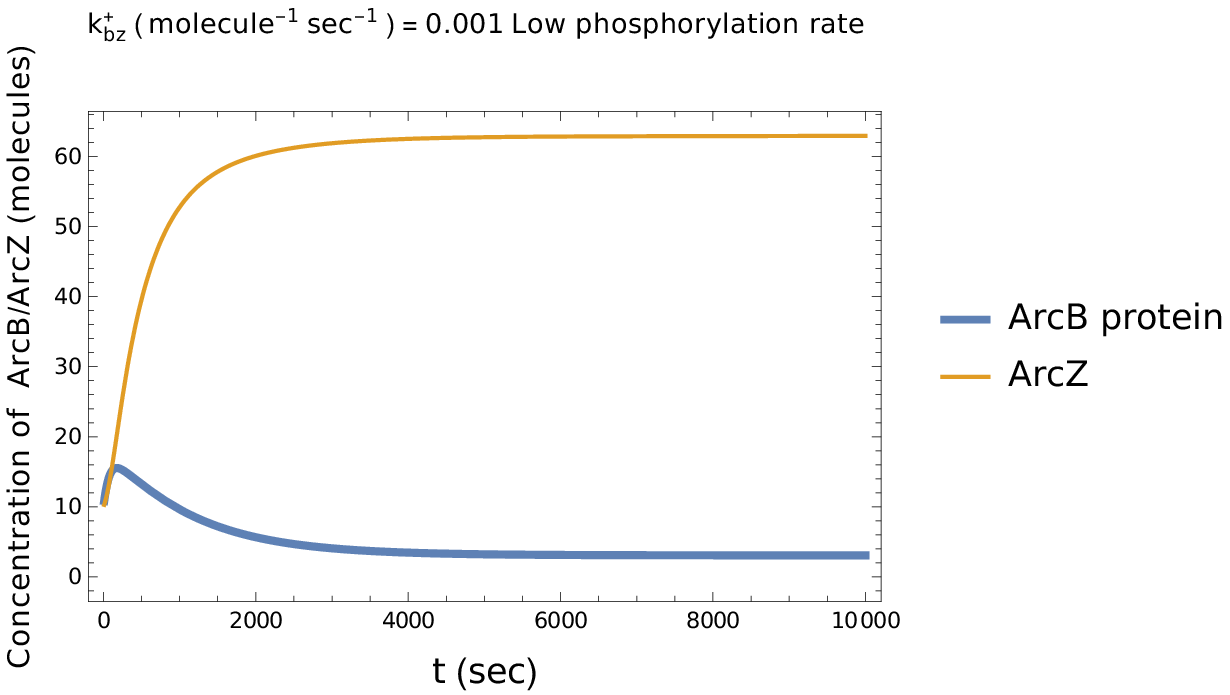}
 \endminipage
 \caption{ The time evolution of the number of ArcB and ArcZ molecules per cell for a high (panel 1) and low (panel 2) phosphorylation rate with 
 $k_{\rm AP}=0.1\ {\rm molecule}^{-1}$ and  $k_{\rm AP}=0.001\  {\rm molecule}^{-1}$, respectively. The remaining parameter values are as mentioned in  table \ref{table-one}. }
\label{fig:high-low-b-z_math}
\end{figure}

\begin{table}[ht]
\caption{Parameter Values}
\centering
\begin{tabular} {|c| c| c| }
\hline\hline
Reaction & Rate Constants & Parameter Values\\
\hline
ArcZ synthesis  & $\beta_z (*)$  & 0.1\ ({\rm molecule/sec})\\
\hline
ArcZ degradation & $\gamma_z$  &0.0025 \ (${\rm sec^{-1}}$) \\
\hline
$\sigma^s$-mRNA synthesis & $\beta_\sigma$ & 0.02 \ ({\rm molecule/sec}) \\
\hline
$\sigma^s$-mRNA degradation & $\gamma_m$  & 0.002\ (${\rm sec^{-1}}$)\\
\hline
ArcB mRNA synthesis & $r_{bm}$  &  0.02 \ ({\rm molecule/sec})\\
\hline
ArcB mRNA degradation &  $\gamma_{bm}$   & 0.002\ (${\rm sec^{-1}}$)\\
\hline
ArcB protein synthesis & $r_{bp}$ & 0.01 \ (${\rm sec^{-1}}$)\\
\hline
ArcB protein degradation & $\gamma_{bp}$ & 0.001\ (${\rm sec^{-1}}$) \\
\hline
ArcB-ArcZ mRNA complex formation & $k_{bz}^+$  & 0.01\  (${\rm molecule^{-1} sec^{-1}}$) \\
\hline
ArcB-ArcZ mRNA complex dissociation (ArcZ mediated degradation of ArcB) & $k_{bz}^-$  & 0.01\ (${\rm sec^{-1}}$) \\
\hline
$\sigma^s$mRNA-ArcZ complex formation   &  $k_{mz}^+$ & 1 \  (${\rm molecule^{-1} sec^{-1}}$) \\
\hline
$\sigma^s$mRNA-ArcZ complex dissociation  & $k_{mz}^-$  & 0.02 \ (${\rm sec^{-1}}$)\\
\hline
Repression of arcZ gene by ArcA-P & $k_z=k_z^+/k_z^-(*)$  & 0.1\ (${\rm molecule^{-1} }$)\\
\hline
Repression of $\sigma^s$ gene by ArcA-P & $k_\sigma=k_\sigma^+/k_\sigma^-(*)$   & 0.1 \ (${\rm molecule^{-1} }$)\\
\hline
Phosphorylation activity & $k_{AP}=\frac{k_p k_c}{1+k_p^+/k_c^-} (*)$ &0.1  {\rm or} 0.001 \ (${\rm molecule^{-1} }$)\\
\hline
\end{tabular}\label{table-one}\\
We have chosen an equilibrium concentration for ArcA molecules, $[A]=60 \ {\rm molecules}$.  The parameter values are obtained 
from  references \cite{shimoni} and \cite{alon}.  For details  on parameters with $*$, see the discussion in the main text. 
\end{table}
 As expected, in case of a high phosphorylation  rate, the transcription of Z-sRNA is tightly repressed and 
this leads to low Z and high B phase. In case of a low phosphorylation rate,  transcriptional repression of Z is low  and this leads to 
a  high Z, low B state.  
Two plots   in figure (\ref{fig:high-low-b-z_math}) show  how the  Z-sRNA and B-mRNA  concentrations evolve with time 
under different phosphorylation rates. 

\subsubsection{Role of sRNA mediated destabilisation of mRNA}
\begin{figure}[!htb]
 \includegraphics[height=0.3\textwidth]{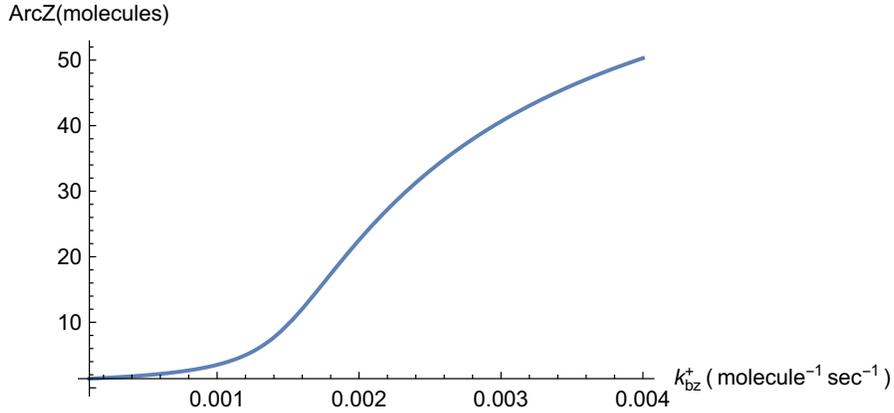}
  \caption{ Number of ArcZ molecules per cell for different values of $k_{\rm bz}^+$  for  a high phosphorylation rate, 
  $k_{AP}=0.1\  {\rm molecule^{-1}}$. The remaining parameter 
  values are as given in table \ref{table-one}. }
\label{fig:arcz-kbzp}
\end{figure}

\begin{figure}[!htb]
 \includegraphics[height=0.3\textwidth]{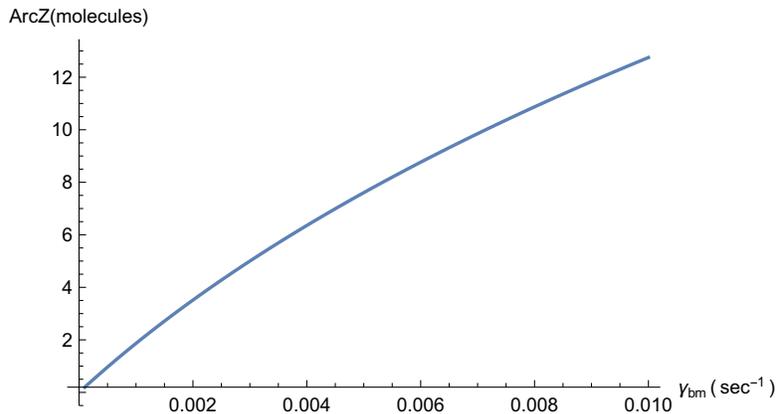}
  \caption{ Number of ArcZ molecules per cell for different values of $\gamma_{bm}$  for  a high phosphorylation rate, $k_{AP}=0.1\ {\rm molecule}^{-1}$. The remaining parameter 
  values are as given in table \ref{table-one}. }
\label{fig:arcz-gambm}
\end{figure}

The dependence of  the equilibrium concentration of Z-sRNA on $k_{bz}^+$ can be understood by analyzing equation (\ref{steadyz}). 
For high phosphorylation rate, that is for a high value of $K$, one may expect $(\gamma_{bm}+K-
\frac{\beta_z}{\gamma_z} G_z^{\rm tot} k_{bz}^+)$ to be large positive. The concentration of Z-sRNA  can be approximated as 
\begin{eqnarray}
[Z]\approx \frac{\gamma_{bm}}{\gamma_z}\beta_z G_z^{\rm tot}/(\gamma_{bm}+K-
\frac{\beta_z}{\gamma_z} G_z^{\rm tot} k_{bz}^+). \label{approxz}
\end{eqnarray}
Thus under  high phosphorylation rate, Z-sRNA  concentration has  a low value which  becomes independent of $k_{bz}^+$ 
for $k_{bz}^+<<(\gamma_{bm}+K)/(\frac{\beta_z}{\gamma_z} G_z^{\rm tot})$ (see figure (\ref{fig:arcz-kbzp})). 
 However, this low concentration scenario  may change drastically  if  the sRNA-mediated mRNA degradation rate ($k_{bz}^+$) 
is increased.  With the increase in this rate, mRNA degradation increases and  this increases the concentration of Z-sRNA 
 abruptly even when 
the phosphorylation rate is high. Equation (\ref{steadyz}) implies that  this abrupt change in concentration must happen when $(\gamma_{bm}+K-
\frac{\beta_z}{\gamma_z} G_z^{\rm tot} k_{bz}^+)$  changes sign with the increase in $k_{bz}^+$. In figure (\ref{fig:arcz-kbzp}), we show how the concentration of 
Z-sRNA increases with $k_{bz}^+$  in case of high phosphorylation rate.  
This drastic behaviour   originating from sRNA-mRNA interaction 
might be compared with threshold linear response seen earlier in the context of sRNA mediated mRNA degradation \cite{hwa,hwa2} even when 
there was no feedback component in the regulation. In the threshold-linear response, the gene expression is completely silenced when 
the target transcription rate is below a threshold. Above the threshold, mRNAs code for the protein leading to a linear increase in the 
protein concentration with the target transcription rate. In case of strong sRNA, mRNA interaction, the transition from one gene expression 
regime to the other is sharp. As the interaction between sRNA and mRNA becomes weaker, the transition becomes smoother although the 
threshold-linear form is preserved. Figure (\ref{fig:arcz-kbzp}) describes a similar smooth transition from a negligible sRNA expression regime to a 
high sRNA  expression regime except for the fact that, beyond the threshold,  the change in the concentration with $k^+_{bz}$  is not linear. 
This entire  phenomenon is the result of   the   nonlinear interactions between 
sRNA and mRNA ( $k_{bz}^+$ dependent term in equation (\ref{diffeqnbm})) 
and  accordingly, no drastic effect, for example,   of an increase in the mRNA degradation rate can be seen  on the sRNA concentration
    (see figure \ref{fig:arcz-gambm}).

\subsection{ Model II}
Applying the steady-state condition on (\ref{diffeqnz1}), (\ref{diffeqnbz}) and (\ref{diffeqnsigmamz}), 
we have the following equation for the steady-state concentration $[Z]$.
\begin{eqnarray}
\frac{\beta_z}{\gamma_z} G_z^{tot}-[Z]= [Z]\frac{{{K}}^2}{k_z(k_{bz}^+ [Z]+\gamma_{bm})^2}
\label{solnz}
\end{eqnarray}
where, as before,  ${K}=\frac{k_z k_{AP} {r_{bp}} r_{bm} [A]}{\gamma_{bp}}$. 

\begin{figure}[ht]
  \includegraphics[height=0.4\textwidth]{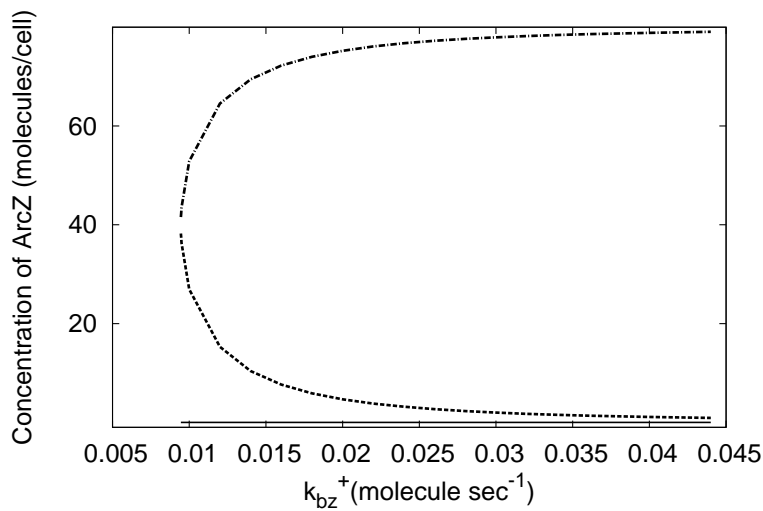}  
\caption{ The three branches of the equilibrium  solution of $[Z]$ for model II. The uppermost (dash-dotted) and the lowermost (solid) lines are the stable branches of the solution. 
The intermediate (dotted) line  corresponds to the unstable branch of the solution.  For this plot $k_{AP}=0.1\ {\rm molecule}^{-1}$. 
The other parameter values are as displayed in table \ref{table-one} except that here $\gamma_{bm}=0.001\  {\rm sec}^{-1}$.  }
\label{fig:bistabilitymain_model2}
\end{figure}

\begin{figure}[ht!]
   \includegraphics[height=0.5\textwidth]{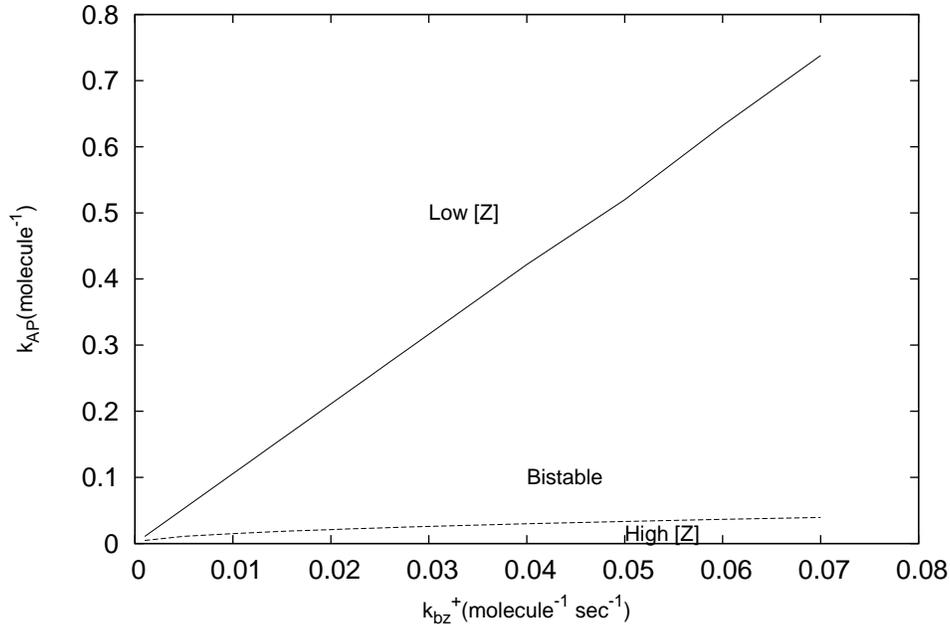}  
\caption{The phase diagram of model II in the $k_{bz}^+$ -$k_{AP}$ parameter space. The 
parameter values are as displayed in table \ref{table-one} except that here $\gamma_{bm}=0.001\  {\rm sec}^{-1}$. }
\label{fig:phasediag2}
\end{figure}

Being a cubic equation in $[Z]$, equation (\ref{solnz})  leads to three equilibrium  solutions for $[Z]$ for certain parameter values. 
Figure (\ref{fig:bistabilitymain_model2}) shows these three solutions over a range of values of $k_{bz}^+$.    The  upper most   
and the lower most  branches of solutions are stable branches, while the intermediate branch is the unstable branch. 
Since in case of model I, the phosphorylation rate and  $k_{bz}^+$ are crucial for the threshold response of   sRNA expression, we are interested 
to know how, for model II,  these two parameters  alter the threshold response behaviour and gives  rise to two possible stable states (a 
bistable response). The steady-state behaviour, in this case, can be 
presented  through a phase diagram in the parameter space of $k_{bz}^+$ and $k_{AP}$. 
 Equation  (\ref{solnz}) can be written as 
 \begin{eqnarray}
 f([Z])=g([Z]), \ \ {\rm where} 
 \end{eqnarray}    
\begin{eqnarray}
f([Z])=\frac{\beta_z}{\gamma_z} G_z^{tot}-[Z] \ \  {\rm and} \ \ 
g([Z])= [Z]\frac{{{K}}^2}{k_z(k_{bz}^+ [Z]+\gamma_{bm})^2}.
\end{eqnarray}
The bistability of $[Z]$ is  found when, for a given parameter values, the linear function $f([Z])$ 
intersects $g([Z])$ at three different $[Z]$ values.   
 The onset of bifurcation (bistability)  happens at  particular parameter values
at which the straight line $f([Z])$  meets $g([Z])$ as a tangent at a particular value of $[Z]$ \cite{strogatz}.  
Thus the  onset of bifurcation can be analysed by solving equation (\ref{solnz}) together with the condition arising from  slope matching  
of the two curves  $f([Z])$ and $g([Z])$ i.e. 
\begin{eqnarray}
\frac{d}{d[Z]} f([Z])=\frac{d}{d[Z]} g([Z]), \label{tangent}
\end{eqnarray}
Solving these two equations for $k_{AP}$ and $[Z]$ for  different values of $k_{bz}^+$   and for fixed values of other parameters, 
we obtain  the bistable  region  in $k_{bz}^+$ and $k_{AP}$ parameter space (see figure (\ref{fig:phasediag2})). 
It is apparent from the phase diagram  that the threshold response that was seen in case of model I, disappears in this case. The threshold response 
was seen for high phosphorylation rate  for which sRNA synthesis is  low  upto a certain  value of $k_{bz}^+$ 
 but increases rapidly as   $k_{bz}^+$ increases beyond this value.  This continuous variation 
 in the value of $[Z]$  from a low to a high value is governed by a single equilibrium solution of $[Z]$ (see figure (\ref{fig:arcz-kbzp})).
 As the phase diagram in figure (\ref{fig:phasediag2}) shows, in case of model II,  for high values  $k_{AP}$, the low $[Z]$ state 
 enters into the bistable region as the  strength of the sRNA-mRNA  interaction,  $k_{bz}^+$,  is increased. 

\subsection{Model III}

\begin{figure}[ht]
  \includegraphics[height=0.4\textwidth]{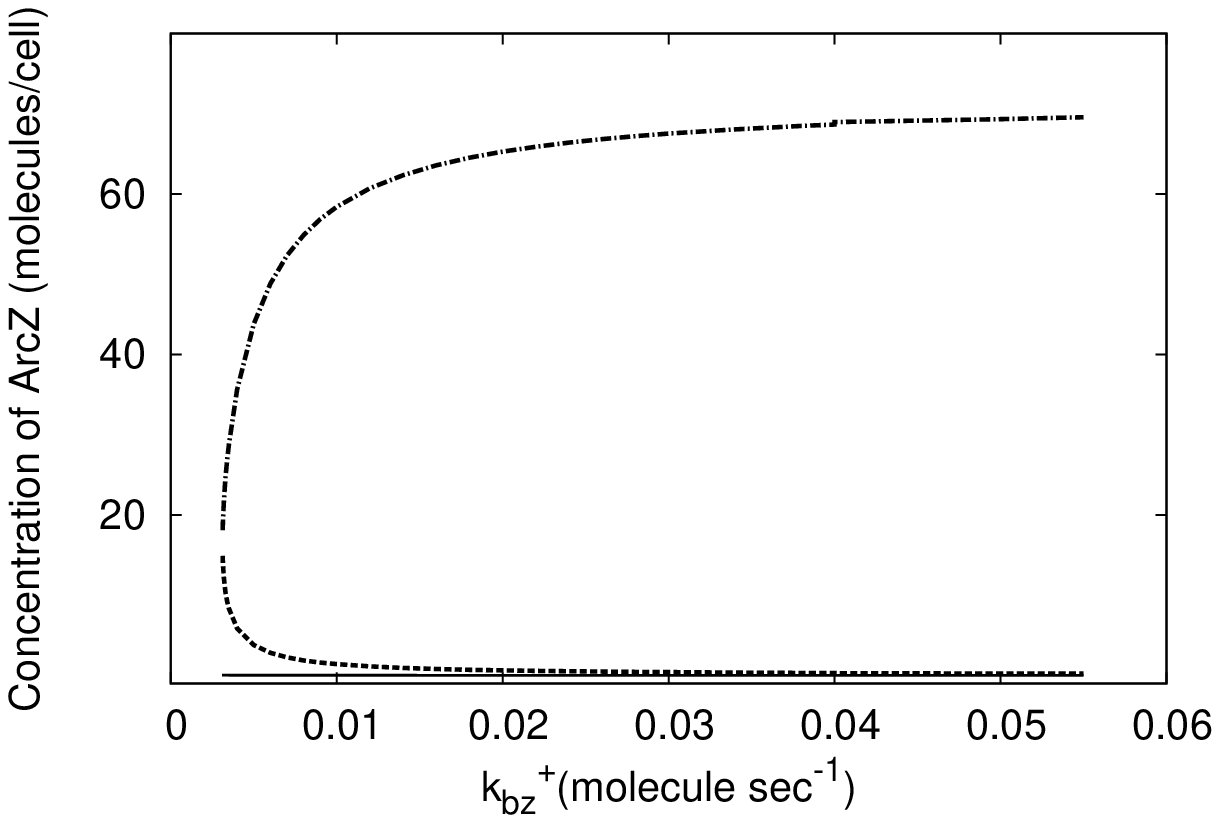}  
\caption{ The three branches of the equilibrium solution of $[Z]$ for model III. The uppermost (dash-dotted) and the 
lowermost (solid) lines are the stable branches of the solution. 
The intermediate (dotted) line  corresponds to the unstable branch of the solution.  For this plot, 
 $k_{AP}=0.1\ {\rm molecule}^{-1}$. 
The other parameter values are as displayed in table 
\ref{table-one} except that here $\gamma_{bm}=0.001\  {\rm sec}^{-1}$.  }
\label{fig:bistabilitymain_model3}
\end{figure}

\begin{figure}[ht!]
   \includegraphics[height=0.5\textwidth]{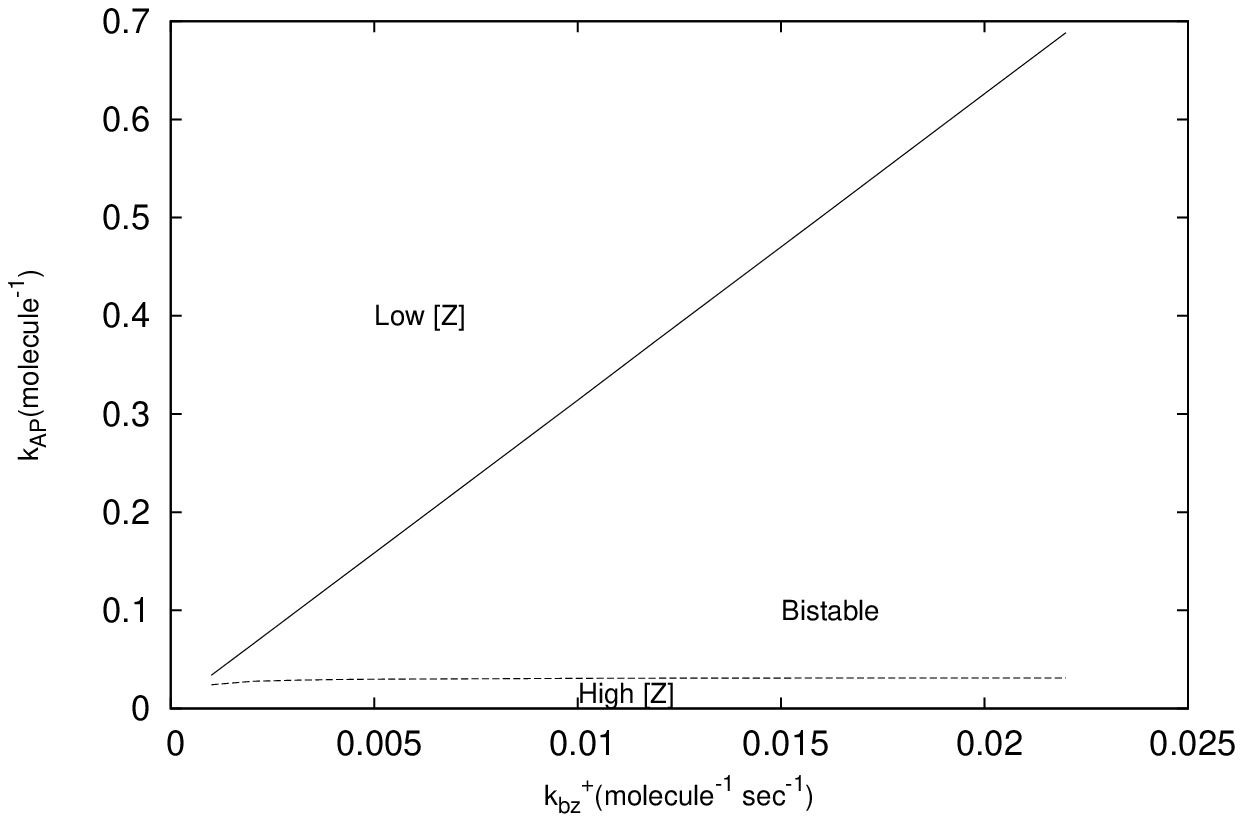}  
\caption{The phase diagram of model III in the $k_{bz}^+$ -$k_{AP}$ parameter space. 
The  parameter values are as displayed in table 
\ref{table-one} except that here $\gamma_{bm}=0.001\  {\rm sec}^{-1}$. }
\label{fig:phasediag3}
\end{figure}

For model III, bistable solutions similar to that of model II are found. As before, the steady-state solutions for $[Z]$ are found from the following equation
\begin{eqnarray}
f([Z])=g([Z]), \ \  {\rm where} \label{bistable3}
\end{eqnarray}
\begin{eqnarray}
f([Z])=\gamma_z(\gamma_{bm}+k_{bz}^+[Z])+k_{bz}^+ r_{bm} \ \ {\rm and}\ \ \ 
g([Z])=\frac{\beta_z  G_z^{\rm tot} (\gamma_{bm}+k_{bz}^+ [Z])^2}{[Z] (\gamma_{bm}+k_{bz}^+[Z]+ K)}\label{fzmodel3}
\end{eqnarray}
The solutions for $[Z]$  as $k_{bz}^+$  is  varied for a specific value of $k_{AP}$ is  shown in figure (\ref{fig:bistabilitymain_model3}).  
Here also,  for a given  value of $k_{bz}^+$,  there is a range of $k_{AP}$ over which bistablity in  $[Z]$ is found.  
The phase diagram obtained upon solving (\ref{bistable3}) and the equation for slope matching ( $\frac{df([Z])}{d[Z]}=\frac{dg([Z])}{d[Z]}$)  
numerically,  is shown in figure (\ref{fig:phasediag3}). 
A comparison with  the phase diagrams of  model II  shows that  model III 
 has a much wider bistable region indicating 
that  the co-degradation of sRNA  gives rise to a more robust 
bistable behaviour in comparison with model II where the bistability arises due to cooperativity. 
  
\section{Stochastic analysis}
\subsection{Methodology}
In our previous analysis, we have framed the time evolution equation for Z-sRNA incorporating various terms originating from 
transcriptional and translational regulation. 
The bistable response found in model II and model III can give rise to  phenotypic heterogeneity in   a 
bacterial population. The differential equation 
method discussed  in the previous section describes the noise-free dynamics of the system and indicates that, in case of model II and model III, 
 there are two stable equilibrium states for a cell with low and high concentration levels of sRNA. 
These two states are separated by an unstable state which sets a barrier  for crossing over 
from one equilibrium state to another. In this picture, the 
equilibrium state of the cell is determined by the initial condition or previous history. For example, if  
cells in a population are in certain initial state supporting 
high or low expression level, all the cells in this population would continue to remain in 
the same state in  the equilibrium   also.  In reality, the gene expression is inherently noisy. 
  Due to the noise,  in experiments, the  signature of 
bistability is found through  the coexistence of  two subpopulations of cells with low and high expression levels \cite{kaern,raj,maamar}.  
This happens due to noise driven 
transitions of a cell from one state to another.  Due to fluctuations (noise) in gene expression, 
a cell in the  low-expression state, for example, may  cross over to 
the high-expression state by overcoming the    barrier corresponding to  the intermediate  
unstable state. Such bistable expression pattern of cells 
has been observed in synthetic networks also \cite{becskei}. The primary aim of  this part of the work 
is to study the effect of noise on the bistable response seen in model II 
and III. Based on the origin, the noise is categorised as  (i) extrinsic  and (ii) intrinsic noise \cite{swain,elowitz}. 
The extrinsic noise is  due to gene independent fluctuations which, for example,  might be due to  fluctuations in 
 the amount of RNA polymerase which itself is a  gene product.  The intrinsic noise, on the other hand, is due to 
gene specific fluctuations which might be, for example, 
due to  random occurrences of various biochemical events associated  with translation and transcription.  
In the past, there have been experimental studies to quantify  the noise and discriminate the two mechanisms
behind the origin of these two types of noise. These studies provide valuable inputs regarding  the role of noise in 
the cell-cell variation in gene expression as well as in intracellular reactions \cite{elowitz}.
 
 In the presence of noise, the concentration levels of various components must be described probabilistically. Here, 
 we use the Fokker-Planck equation based method  to obtain the probability distributions of sRNA concentrations in various cases. 
The bistable response, observed using  the differential equation  method, leads to a bimodal probability distribution in the Fokker-Planck approach. 
Such bimodal distributions indicate that there are significant  probabilities for a cell to be in two distinct states with distinctly different protein concentration levels. 
Further, this analysis allows us to find how the  noise strength or the noise correlation influences the nature of the distribution. 
 The present section is primarily devoted to  general,  formulation related discussions. Model specific results are derived in the following subsections.  

The stochastic analysis begins with the Langevin equation which  describes the time evolution of the sRNA concentration 
  as shown in  equation  (\ref{diffeqnz1}) or (\ref{diffeqnz2}) along with additional noise terms \cite{hasty,sayantari,zheng}.  In
   order to simplify notations, we denote the concentration of Z-sRNA as $z$ below.
  Considering  the presence of two different types of noise, we write below the Langevin equation with both  multiplicative and additive noise
    \begin{eqnarray}
\frac{dz}{dt}=f(z)+g(z)\epsilon(t)+\eta(t).\label{langevin}
\end{eqnarray}
The forms of $f(z)$ and $g(z)$  change depending on the model considered.  
As discussed in detail in  the next subsection,  for model II,  the explicit forms of $f(z)$ and $g(z)$ 
can be obtained  by separating the deterministic and the noisy parts of the first term in 
equation (\ref{diffeqnz1}). In a similar way, for model III,  the forms of $f(z)$ and $g(z)$  are obtained from equation (\ref{diffeqnz2})  (see subsection (\ref{ssec:stochmodel3})).
The additive noise, denoted by $\eta(t)$,  represents  random fluctuations in the external environment and it is usually not directly related to the 
gene concerned. The multiplicative noise, $\epsilon(t)$, on the other hand, 
represents the intrinsic noise that originates from the   randomness in various  biochemical events  associated with the transcription or 
translation.  Such transcription or translation  processes, for example,  are regulated by the participation of various regulatory molecules 
whose presence or absence is somewhat random  leading to random rate parameters. The effects of all such 
randomness are lumped together into a single multiplicative noise term $g(z) \epsilon(t)$. 
The  multiplicative and additive noise obey Gaussian distribution with  the following correlations
\begin{eqnarray}
\langle \epsilon(t)\epsilon(t') \rangle = 2D_1\delta(t - t'),\\
\langle\eta(t)\eta(t') \rangle = 2D_2\delta(t -t') \ \ \ {\rm and}\\
\langle\eta(t)\epsilon(t') \rangle = \langle \epsilon(t)\eta(t') \rangle= 2\lambda \sqrt{D_1D_2}\delta(t - t').
\end{eqnarray}
Here $D_1$ and $D_2$ denote the strength of the two types of noise $\epsilon(t)$ and $\eta(t)$, respectively.  $\lambda$ denotes the correlation 
between the  additive and the multiplicative noise. For $\lambda=0$, the two types of noise are uncorrelated. $\epsilon(t)$ and 
$\eta(t)$ are expected to be correlated ($\lambda\neq 0$)  when these two types of noise  are  of same origin.  
$f(z)$ in equation (\ref{langevin})  takes into account the deterministic parts in equation (\ref{diffeqnz1}) or (\ref{diffeqnz2}). 

Beginning with this Langevin equation, one may obtain the Fokker-Planck (FP) equation following the 
 standard  procedure \cite{reichl}. The FP equation describes the time evolution of  the probability density $\rho(z,t)$ 
 with   $\rho(z,t) dz$ denoting the probability of having the concentration, $z$,  between $z$ and $z+dz$ at time $t$.  
 The FP equation derived from (\ref{langevin}) is 
 \begin{eqnarray}
\frac{d\rho}{dt}=-\frac{d}{dz}[A(z) \rho(z,t)]+\frac{d^2}{dz^2} (B(z) \rho(z,t)),
\end{eqnarray}
where 
\begin{eqnarray}
&& A(z)=f(z)+D_1 g(z){\big(}\frac{d}{dz} g(z){\big)}+\sqrt{D_1 D_2}\lambda {\big(}\frac{d}{dz}g(z){\big)}, \  \ {\rm and}\ \ \\
&& B(z)=D_1 g(z)^2 +2\lambda \sqrt{D_1 D_2} g(z)+D_2.
\end{eqnarray}
In the steady-state,  the probability distribution is independent of time ($\frac{d\rho(z,t)}{dt}=0$). The solution for the 
  steady-state  probability density is  
\begin{eqnarray}
\rho(z)=\frac{N}{B(z)} \exp[\int^z \frac{A(z')}{B(z')} dz' ],
\end{eqnarray}
where $N$ is the normalisation factor. 
$\rho(z)$  can be further simplified as 
\begin{eqnarray}
 \rho(z)=\frac{N}{\{D_1 g(z)^2 +2\lambda \sqrt{D_1 D_2} g(z)+D_2\}^{1/2}} \exp[\int^z dz'\  \frac{f(z')}{\{D_1 g(z')^2 +2\lambda \sqrt{D_1 D_2} g(z')+D_2\}}.
 \label{finaldist}
 \end{eqnarray}
 Following this general formulation, we obtain the probability distributions for specific models below. 
\subsection{Model II}
\label{ssec:stochmodel2}

For model II, we proceed with equation (\ref{langevin}) with 
\begin{eqnarray}
f(z)=p_0\frac{(z+\gamma)^2}{(z+\gamma)^2+p_2}-\gamma_z z \ \ {\rm and}\ \  g(z)=\frac{(z+\gamma)^2}{(z+\gamma)^2+p_2}
\label{fzgz2}
\end{eqnarray} These forms of $f(z)$ and $g(z)$ are  obtained by using the expressions for steady concentrations of $[AP]$, $[B_p]$ and $[B_m]$ 
in equation (\ref{diffeqnz1}). We have assumed 
steady-state conditions, $\frac{d}{dt}[\sigma^smZ]=\frac{d}{dt}[B_mZ]=0$  (see equations (\ref{diffeqnbz}) and (\ref{diffeqnsigmamz}))
for the  concentrations of mRNA-sRNA complexes, $[\sigma^smZ]$ and $[B_m Z]$.
In equation (\ref{fzgz2}), $p_0=\beta_z G_z^{\rm tot}$, $p_2=\frac{K^2}{k_z {k_{bz}^+}^2}$ and $\gamma=\frac{\gamma_{bm}}{k_{bz}^+}$. 
The form of $g(z)$ in (\ref{fzgz2}) can be understood  from the origin of the  multiplicative noise. Here we assume that various biochemical events 
associated with  transcription  makes  $p_0$, effectively,  noisy. Thus  a  noise in the form  $p_0\rightarrow p_0+\epsilon(t)$ leads to the above expression 
of $g(z)$. Substituting these forms of $f(z)$ and $g(z)$ in equation (\ref{finaldist}), we find the probability distribution numerically 
 for different cases such as only additive noise, additive and multiplicative noise without correlation ($\lambda=0$)
 and finally,  additive and multiplicative noise with correlation ($\lambda\neq 0$). 

\begin{figure}[ht!]
\begin{tabular}{cc}
\subfloat[$p_0=0.13$]{\includegraphics[width = 4.5in]{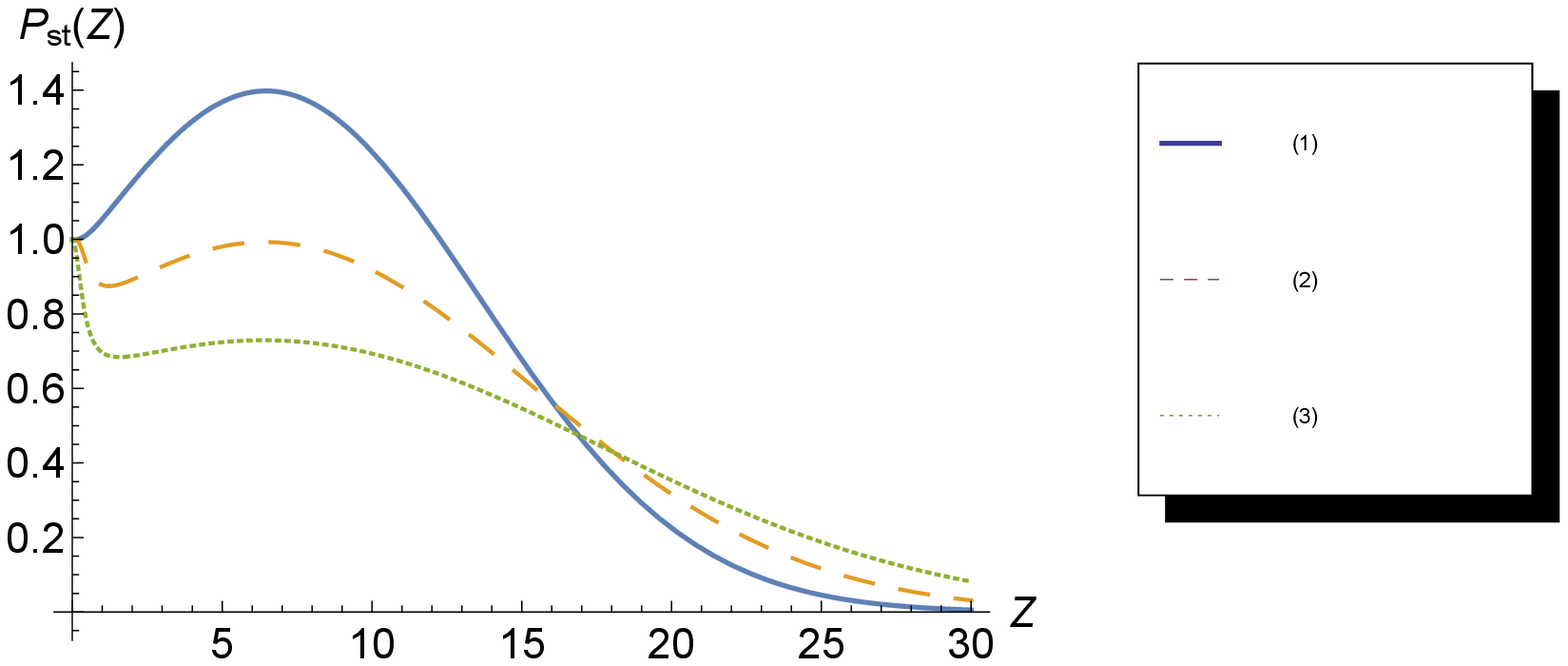}}&
\subfloat[$p_0=0.13$]{\includegraphics[width = 2in]{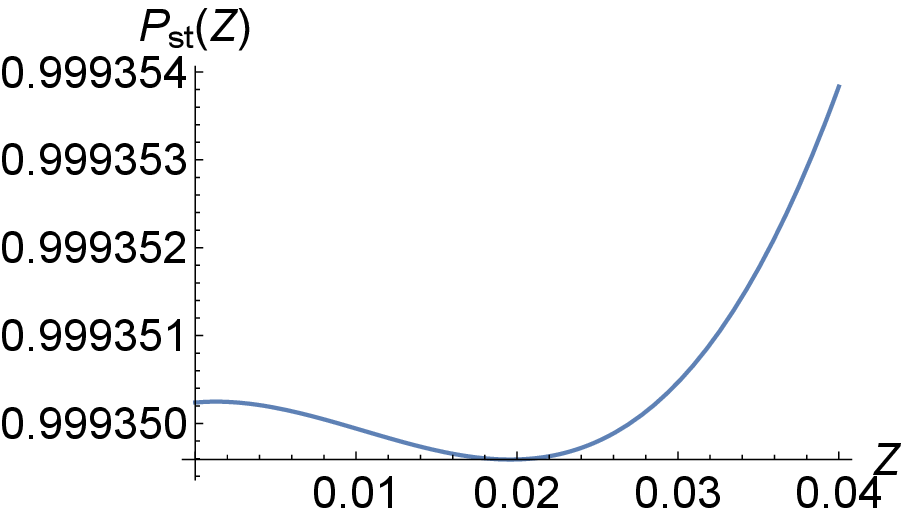}} 
\end{tabular}\\
\begin{tabular}{cc}
\subfloat[$p_0=0.09$ ]{\includegraphics[width = 4.5in]{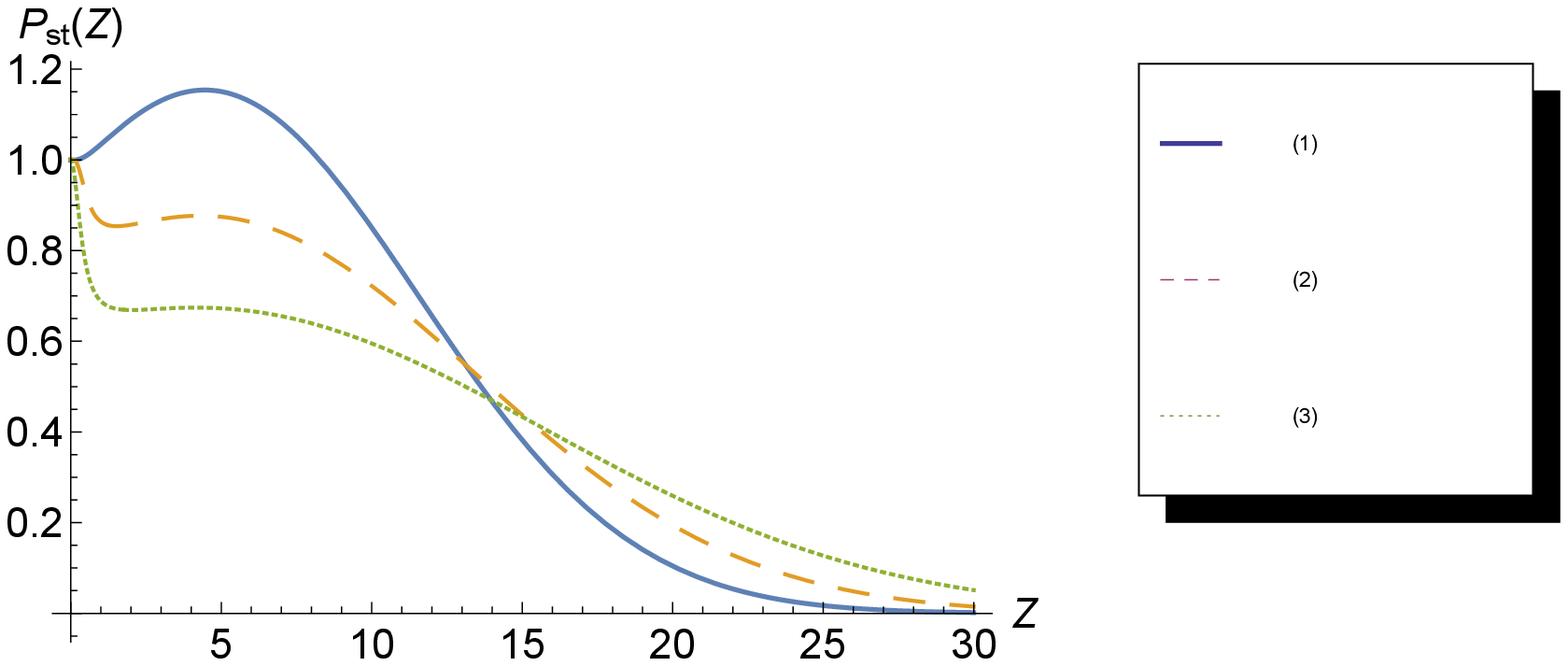}}
\subfloat[$p_0=0.09$ ]{\includegraphics[width = 2in]{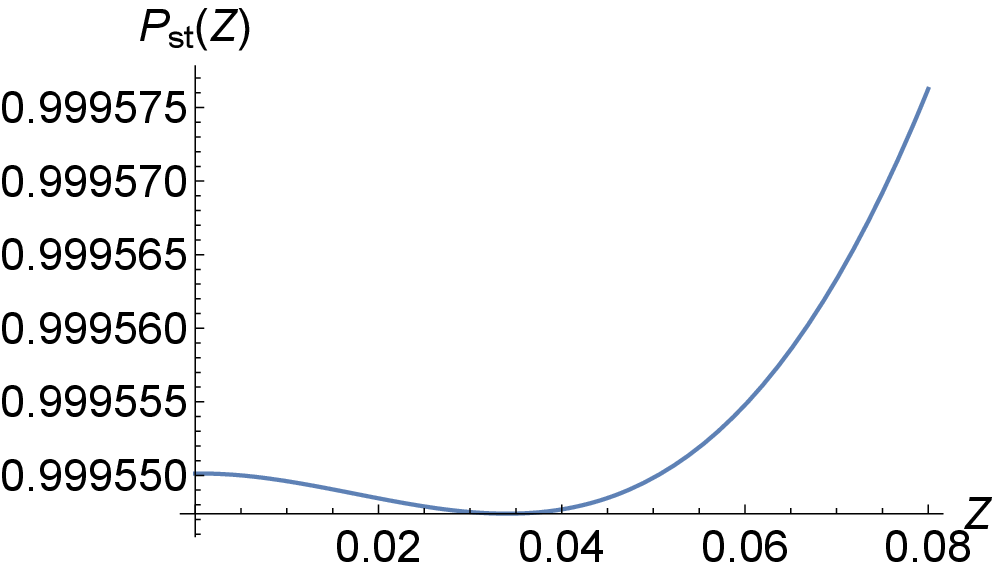}}
 \end{tabular}
 \caption{Probability distributions for model II with  (1) additive noise ($D_2=1$) (2) additive and multiplicative noise without correlation  ($D_2=1$, $D_1=0.6$ and $\lambda=0$)
(3) additive and multiplicative noise with correlations ($D_2=1$, $D_1=0.6$  and $\lambda=0.6$)
are plotted. 
Other parameter values are $\gamma=0.005$, $p_2=0.2$ and $\gamma_z=0.02$. Since the peak in the probability distributions near low $z$ is not seen 
in  figures (a) and (c), we have shown   in figures 
(b) and (d) the respective  distributions near the low concentration region with appropriate zooming.}
\label{fig:alldist_transcrip_homodimer}
 \end{figure}

 The probability distributions for the  concentration, $z$, are bimodal with two peaks near high and low $z$ values. 
 As shown  in figure (\ref{fig:alldist_transcrip_homodimer}), in case of additive noise,  the  peak
 in the distribution at low  $z$ value    is much smaller in comparison to the peak  at high $z$ value. 
 This indicates that the high z  state is more favorable in comparison to low z states.  This scenario, however, changes  since with the increase in the 
  noise strength and the correlation between the multiplicative and additive noise, the  peak of the distribution at low z value becomes significantly 
  more prominent. 
   This implies that  the cells with low concentration of sRNA are more probable as the strength of the multiplicative noise  
  or the correlation between the two types of noise is increased. 
 Further, the probability distributions become broader as the strength of the noise and the  correlation between the additive and multiplicative noise increases. 
As we  show below,  this feature is  not observed for model III.   Such broadening of  
 distributions indicates  that the noise strength or the noise correlation   might give rise to noise driven heterogeneity 
 in the microbial population. 

\subsection{\label{ssec:stochmodel3} Model III}
For model III,
\begin{eqnarray}
f(z)=\frac{p_0(z+\gamma)}{(z+p_1)}-r_{bm} \frac{z}{z+\gamma}- \gamma_z z \label{fz}.
\end{eqnarray}
These forms  are again  obtained by using the expressions for steady concentrations of $[AP]$, $[B_p]$ and $[B_m]$ in equation (\ref{diffeqnz2}).
For the  concentrations of mRNA-sRNA complex, $[\sigma^smZ]$, we have assumed 
a steady-state condition, $\frac{d}{dt}[\sigma^smZ]=0$  (see equation (\ref{diffeqnsigmamz})).
In (\ref{fz}), $p_0=\beta_z G^z_{\rm tot}$, $\gamma=\frac{\gamma_{bm}}{k_{bz}^+}$ and 
$p_1=\gamma+\frac{K}{k_{bz}^+ }$. 
As before, the form of $g(z)$ can be understood  from the origin of the  multiplicative noise. Here, we consider
 two possibilities for  the multiplicative noise. If the  fluctuations in various rate parameters 
 effectively lead to a noisy $p_0$,  we may choose    $p_0\rightarrow p_0+\epsilon(t)$ such that 
$g(z)=\frac{z+\gamma}{z+p_1}$.
The other possibility that we consider here is an effective fluctuation or noise 
 in   $r_{bm}$ as $r_{bm}\rightarrow r_{bm}+\epsilon(t)$ 
for which $g(z)=-\frac{z}{z+\gamma}$. 
 
\begin{figure}[ht!]
\begin{tabular}{cc}
\subfloat[$p_0=0.26$]{\includegraphics[width = 2.5in]{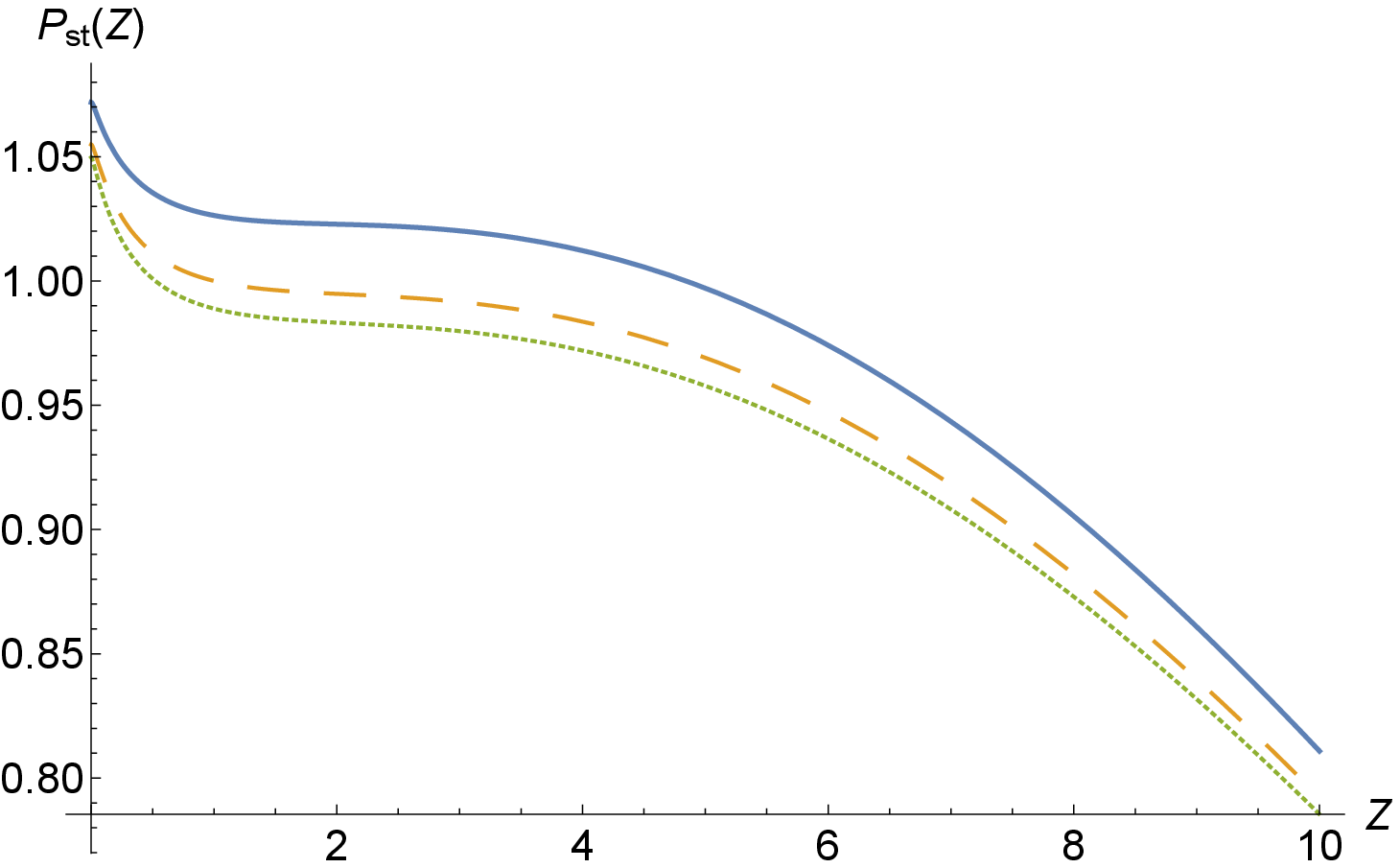}} &
\subfloat[$p_0=0.28$]{\includegraphics[width = 2.5in]{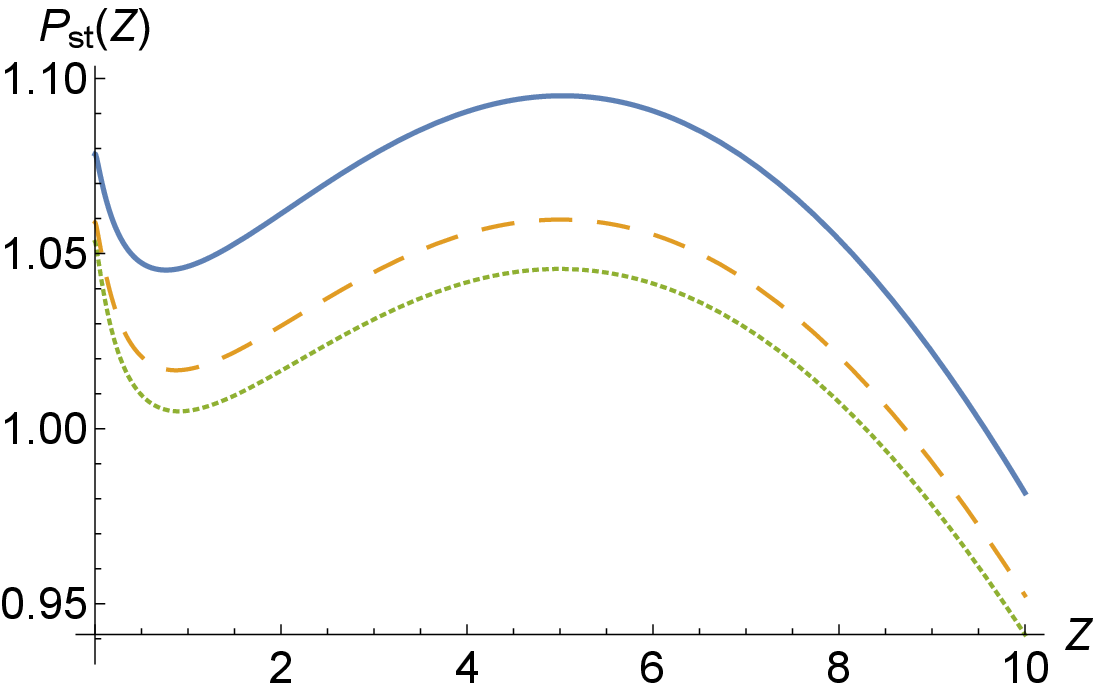}} 
\end{tabular}\\
\begin{tabular}{c}
\subfloat[$p_0=0.32$ ]{\includegraphics[width = 6in]{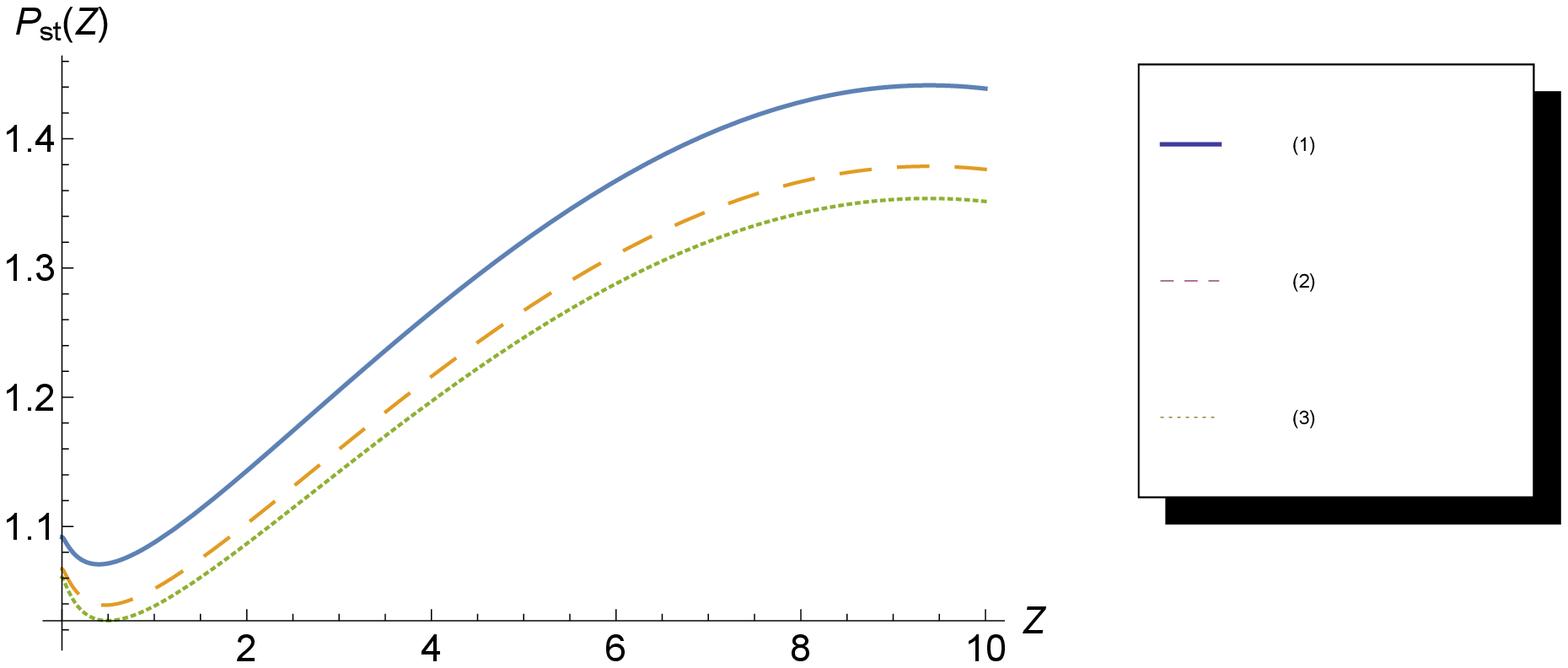}}
 \end{tabular}
\caption{Probability distributions for model III with  $g(z)=\frac{z+\gamma}{z+p_1}$  and 
with  (1) additive noise ($D_2=1$, $D_1=0$ and $\lambda=0$) (2) additive and multiplicative noise without correlation   ($D_2=1$, 
$D_1=0.03$ and $\lambda=0$)
(3) additive and multiplicative noise with correlations ($D_2=1$, $D_1=0.03$  and $\lambda=0.05$)
are plotted. Other parameter values are $\gamma=0.01$, $p_1=0.2$, $r_{bm}=0.22$ and $\gamma_z=0.01$. }
\label{fig:alldist_transcrip_codegrade}
\end{figure}

 Using  these  forms of $f(z)$ and $g(z)$, we obtain   stationary distributions   for only additive noise, both additive and 
 multiplicative noise without correlation ($\lambda=0$) and 
 additive and multiplicative noise with correlation ($\lambda\neq 0$.).  The 
 distributions for $g(z)=\frac{z+\gamma}{z+p_1}$  are displayed in  figure (\ref{fig:alldist_transcrip_codegrade}). 
    The  figure shows the role of the synthesis rate of sRNA, i.e. $p_0$,
 on the  probability distributions under different types of noise.  
The probability distribution  for $p_0=0.28$  is  prominently bimodal with two  peaks at low  and high $z$ values. 
For synthesis rates lower than  $p_0=0.28$, the  peak at large $z$  becomes less prominent indicating 
that  the low sRNA  expression state is more favourable. 
For high synthesis rate, the peak at high $z$   becomes more  prominent showing a tendency of  having  higher concentration of  sRNA. 
An additional figure presented in appendix (\ref{noise}) (see figure 
(\ref{fig:probdist_compare_transcrip_codegrade}) in the appendix) shows that, for a given synthesis rate,   as the noise strength or the 
correlation between the two types of noise 
 increases, the peak in the probability distribution  near the low  $z$ value    becomes more prominent  indicating that the 
 cells are more likely to be in the low sRNA  expression state as compared to the high expression state. 
Figure (\ref{fig:alldist_translation_codegrade}) in appendix (\ref{noise}) shows  the 
probability distributions  for $g(z)=-\frac{z}{z+\gamma}$ for different synthesis rates of Z-sRNA. 
The role of correlation ($\lambda$) between the two different types of noise  appears to be different 
from the previous case (Figure (\ref{fig:alldist_transcrip_codegrade})).  
 
Finally,   for all these cases with different  noise strength  and noise correlations, no trend of broadening of the probability distribution is found. 
This feature is in sharp contrast to model II, where the probability distribution  broadens with the increase in the noise strength and noise correlations.  
  \pagebreak

\section{Discussion}
In this paper, we have shown that  subtle changes in  the architecture of a minimal network 
involving transcriptional and translational regulation  can bring in drastic changes in the end result. 
This analysis is done on an Arc protein network responsible for  regulation in the 
synthesis of $\sigma^s$ in {\it E.coli} under stress due  to  oxygen and energy availability.  We are particularly interested in a  network motif 
that uses dual strategies involving both transcriptional and translational regulation. The histidine sensor kinase ArcB phosphorylates a protein regulator 
ArcA which  represses  transcription of  ArcZ sRNA. ArcZ sRNA, through a feedback  mechanism,  destabilises ArcB mRNA and thus 
autoregulates  its own synthesis. Starting with 
this network, we have considered three different  scenarios. While in the first case (Model I), the sRNA synthesis is 
repressed  by a single  repressor molecule of phosphorylated ArcA, in the second case (Model II),  the synthesis of sRNA is  repressed by a homodimer 
of the repressor molecule.  In  the third case (Model III), we introduce co-degradation  of the 
sRNA along with the target  mRNA.  In all these  cases, the networks show ultrasensitive response  in the 
sRNA  concentration. We show that while in case of Model I, the concentration of sRNA shows a monostable response with a 
threshold behaviour, 
in case of Model II and Model III,  the sRNA concentration shows a bistable response.  Intuitively, as  per  the 
network in figure (\ref{fig:samplefig1}), 
 for high phosphorylation rates,  the synthesis of sRNA is tightly repressed. 
However, as the sRNA mediated mRNA degradation rate is enhanced,  the sRNA  concentration increases drastically beyond 
 a threshold value of this rate. In general, such threshold behavior is believed to be necessary for controlling cellular response 
 which might be expensive in terms of energy or nutrient. Further, through this mechanism, an immediate response to transient 
 signals can also  be avoided. 
The threshold response, however,  disappears  completely  in case of model II and III.  
In these cases,  at a high phosphorylation rate and for low  rate of sRNA mediated mRNA degradation,  
the sRNA concentration is low. However, as the sRNA mediated mRNA degradation  rate is increased, the 
sRNA concentration shows a  bistable response with high and low values of the 
 concentration. Further,   the phase diagram, in terms of the  phosphorylation rate and the sRNA mediated mRNA degradation 
 rate,   obtained for model II and model III  shows that the   bistable response of     model III 
is  more   robust with a much wider bistable region as compared to model II.  

The noise in gene expression is known to give rise to phenotypic transitions from one state to another  \cite{maamar}. 
For example, in case of bistability, there might be  large fluctuations (noise) due to which a cell switches from one stable 
state to another  upon crossing  the barrier caused by the intermediate unstable state.  
Such noise driven transitions may  lead to two distinct subpopulations in a bacterial colony  and, thereby,   enhance
the  survival probability of bacteria under stress since all the cells  do not suffer the same fate.
In the next part, we have studied the influence of noise on the bistable response found in the  deterministic analysis of model II and model III.  
 Broadly, the gene expression is  affected   by two types
of noise (i) extrinsic  and (ii) intrinsic noise. While the extrinsic noise is due to the perturbation in the external 
environment  unrelated to the gene(s)  involved, the 
intrinsic noise arises, for example, due to various biochemical processes associated with the 
translation or transcription of the gene involved.   As an example of the latter, although 
transcription is often considered as a single biochemical process,  it is actually preceded by a sequence 
of biochemical  events whose random occurrences  might effectively introduce noise  in the transcription 
rate.  Considering that the  intrinsic noise 
 in a specific cellular component can be a source of extrinsic noise for other cellular components, we have 
 proceeded with a  general formulation by including  both extrinsic and intrinsic noise terms.  
 The extrinsic and intrinsic noise are incorporated   through  additive and multiplicative noise terms in the Langevin equations
 appropriate for model II and model III.
Beginning with such  Langevin equations, we  obtain the Fokker-Planck equations which  describe the time 
evolution of the probability distribution of  sRNA concentration for the two models.
  The steady-state probability distributions for both models  are  found to be bimodal with two peaks at low and high  sRNA concentrations.
  The peaks correspond to the high and low expression stable states  found  in the deterministic analysis    of model II and III showing bistability.
 The stochastic formulation allows us to study 
how the  probability distributions especially the peaks in the distributions  change 
as the noise strength  and the  noise correlation are altered. 
In general, the stochastic  analysis shows  that as the noise strength  or the correlation between  the two types of 
 noise is increased, the  peak in the probability distribution 
 near the low sRNA concentration becomes more pronounced indicating 
  an increased possibility of finding cells with  low sRNA concentration in an ensemble of cells.  
We also observe an interesting feature    that 
  unlike model III, model II shows a broadening of the   probability distributions with the  noise strength  
  or the noise correlation.  Such  broadening of the probability distribution 
 hints towards an increased  population heterogeneity in the bacterial colony. 
 
Although the work is motivated from a specific network motif  relevant for bacterial stress response, 
various models  discussed here might, in general,  have  broad  implications  due to  their drastically 
different  outcomes. Further,  the models and the   results might be of relevance in the context of  
designing  synthetic networks leading to  artificial control. 

  \appendix
\section{Reaction Scheme}
\label{reaction}
 Different biochemical reactions that we have considered  are 
    \begin{eqnarray}
  && ArcA+ArcB \rightarrow ArcA \mbox{-}ArcB\  {\rm (ArcA\  and\  ArcB\  complex\  formation)}\\
  && ArcA\mbox{-}ArcB \rightarrow ArcA\mbox{-}P+ArcB \ {\rm( Phosphorylation\  of\  ArcA)}\\
 && ArcA\mbox{-}ArcB\rightarrow ArcA+ArcB\  {\rm (ArcA\mbox{-}ArcB\ complex\ dissociation)}\\
  && ArcA\mbox{-}P\rightarrow ArcA+ P\\
 &&  ArcA\mbox{-}P+G_\sigma\rightarrow \overline{G_\sigma}\  {\rm (transcriptional\ repression\  by ArcA\mbox{-}P)}\\
&&  \overline{G_\sigma}\rightarrow G_\sigma+ArcA\mbox{-}P\\
&&   G_\sigma\rightarrow \sigma^sm+G_\sigma\\
&&   ArcA\mbox{-}P+G_z\rightarrow \overline{G_z}\  {\rm (transcriptional\ repression\  by ArcA\mbox{-}P)}\\
&&   \overline{G_z}\rightarrow G_z+ArcA\mbox{-}P\\
&&   G_z\rightarrow ArcZ+G_z\ {\rm (synthesis\ of\  small\ regulatory\ RNA,\  ArcZ)}\\
 &&  ArcZ+\sigma^sm\rightarrow \sigma^sm\mbox{-}ArcZ\ {\rm (binding\ of\ ArcZ\ and\ \sigma^s\ mRNA)}\\
 &&  \sigma^sm\mbox{-}ArcZ\rightarrow \sigma^sm+ArcZ\\
 &&  \sigma^sm\mbox{-}ArcZ\rightarrow \sigma^s+\sigma^sm\mbox{-}ArcZ\  {\rm (synthesis\ of\  \sigma^s)}\\
 &&  ArcZ+ArcBm\rightarrow ArcBm\mbox{-}ArcZ\ {\rm (complex\  of\  ArcB\  and\  ArcZ)}\\
 &&  ArcBm\mbox{-}ArcZ\rightarrow ArcZ \ {\rm (ArcZ\  mediated\  degradation\  of\  ArcB\ mRNA)}\\
 &&  \sigma^sm\rightarrow \emptyset \ ({\rm degradation\  of\  \sigma^s\  mRNA})\\
 &&  \emptyset \rightarrow ArcA \ {\rm (supply\  of\  ArcA)}\\
 &&  \emptyset \rightarrow ArcBm\ {\rm (supply\  of\  ArcB\ mRNA)}\\
  &&  \emptyset \rightarrow ArcB\ {\rm (synthesis \  of\  ArcB \ protein)}\\
 &&  ArcA\rightarrow \emptyset\  {\rm (degradation\  of\  ArcA\  protein)}\\
 &&  ArcBm\rightarrow \emptyset \ {\rm (degradation\  of\  ArcB\  mRNA)}\\
 &&  ArcB\rightarrow \emptyset \ {\rm (degradation\  of\  ArcB\  protein)}\\
 &&  ArcZ\rightarrow \emptyset \ {\rm (degradation\  of\  ArcZ)}
  \end{eqnarray}
 Here $\sigma^sm$, ArcBm, represent  $\sigma^s$ mRNA, ArcB mRNA, respectively. 
 ArcZ, ArcA  represent the small regulatory RNA ArcZ and ArcA protein, respectively. 
Further,  ArcA-ArcB, ArcB-ArcZ and ArcA-P represent  
ArcA-ArcB complex, ArcB-ArcZ complex, a  phosphorylated 
ArcA molecule,  respectively. $G_\sigma$ and $G_z$ represent average number of 
$\sigma^s$  and {\it arcz} genes,  respectively. 
$\overline{G_\sigma}$ and $\overline{G_z}$ represent the inactive states of the 
genes upon the action of the transcriptional repressor ArcA-P.  
$\sigma^sm\mbox{-}$ArcZ and ArcBm$\mbox{-}$ArcZ represent the 
bound complexes  of $\sigma^s$ mRNA and ArcZ,   ArcB mRNA and ArcZ, respectively.

\section{Phosphorylation of ArcA }
\label{const-A}
\begin{eqnarray}
ArcA+ArcB\underset{k_c^-} {\stackrel{k_c^+}{\rightleftharpoons}} ArcAB
{\stackrel{k_p^+}\rightarrow} ArcAP+ArcB\\
ArcAP{\stackrel {k_p^-}\rightarrow }ArcA+P 
\end{eqnarray}
These phosphorylation reactions are described by the following differential equations
\begin{eqnarray}
\frac{d[A]}{dt}=-k_c^+ [A][B_p]+k_c^- [AB_p]+k_p^- [AP]+r_A-\gamma_A [A]\label{diffeqna}\\
\frac{d[AP]}{dt}=k_p^+ [AB_p]-k_p^-[AP]\label{diffeqnap}\\
\frac{d[AB_p]}{dt} =k_c^+[A][B_p]-k_c^-[AB_p]-k_p^+ [AB_p]\label{diffeqnab}.
\end{eqnarray}
As was mentioned in the main text,  $[A]$, $[AP]$, $[AB_p]$ denote the concentrations of ArcA, phosphorylated   ArcA  and ArcA, ArcB protein
 complexes, respectively. 
 Possible nontrivial equilibrium solutions of (\ref{diffeqna}), (\ref{diffeqnap}) and (\ref{diffeqnab})  are 
\begin{eqnarray}
&&[A]=\frac{r_A}{\gamma_A}, \ \ [AP]=\frac{k_p^+}{k_p^-} [AB_p],  \label{steadyap}\\
&& [AB_p]=\frac{k_c^+}{k_c^-+k_p^+} [A][B_p].\label{steadyab}
\end{eqnarray}
Combining   equations  (\ref{steadyap}) and (\ref{steadyab}), we find 
\begin{eqnarray}
[AP]=\frac{k_p^+}{k_p^-} \frac{k_c^+}{k_c^-+k_p^+} [A] [B_p]=\frac{k_p k_c}{1+k_p^+/k_c^-} [A][B_p]=k_{AP} [A] [B_p]
\end{eqnarray}
where $k_p=\frac{k_p^+}{k_p^-}$, $k_c=\frac{k_c^+}{k_c^-}$ and $k_{AP}=\frac{k_p k_c}{1+k_p^+/k_c^-}$.

For all calculations, we have assumed a fixed  value for the  steady concentration of  ArcA.

 \section{Repressor activity of phosphorylated ArcA}
 \label{synthesis}
Let $G_\sigma$, $\bar{G_\sigma}$ represent the average number  of 
the repressor-free and repressor-bound forms 
of the $\sigma^s$ gene, respectively. Similar meaning is implied here for  $G_z$ and   $\bar{G_z}$ for the ArcZ expressing 
gene. Let us assume 
\begin{eqnarray}
G_\sigma+\bar{G_\sigma}=G_\sigma^{\rm tot}\label{gsigmatotal}\\
G_z+\bar{G_z}=G_z^{\rm tot}.\label{gztotal}
\end{eqnarray}
With the reaction scheme as 
\begin{eqnarray}
G_\sigma+ArcAP \underset{k_\sigma^-} {\stackrel{k_\sigma^+}{\rightleftharpoons}} \bar{G_\sigma}\ \ {\rm and} \ \ 
G_z+ArcAP \underset{k_z^-} {\stackrel{k_z^+}{\rightleftharpoons}} \bar{G_z}, \label{repression}
\end{eqnarray}
the time evolution of these average numbers can be described through equations similar to 
\begin{eqnarray}
\frac{dG_\sigma}{dt}=k_\sigma^- \bar{G_\sigma}-k_\sigma^+ G_\sigma [AP].\label{diffeqngsigma}
\end{eqnarray}
Employing  equilibrium  condition on (\ref{diffeqngsigma}) and using (\ref{gsigmatotal}), 
 we have 
\begin{eqnarray}
G_\sigma=\frac{G_\sigma^{\rm tot}}{1+k_\sigma[AP]}\ \  {\rm and}\ \ 
G_z=\frac{G_z^{\rm tot}}{1+k_z [AP]},
\end{eqnarray}
where $k_\sigma= \frac{k_\sigma^+}{k_\sigma^-}$ and 
 $k_z=\frac{k_z^+}{k_z^-}$.  There  are 
 other loss and gain terms in equation (\ref{diffeqnap}) due to the repression activities  of [AP]. However, these terms will
 not contribute once the equilibrium conditions  as used in equation (\ref{diffeqngsigma})  
 are imposed. 
 
 \section{The role of noise on the peaks in the distribution}
 \label{noise}
\begin{figure}[ht!]
\begin{tabular}{ccc}
{\includegraphics[width = 6.5in]{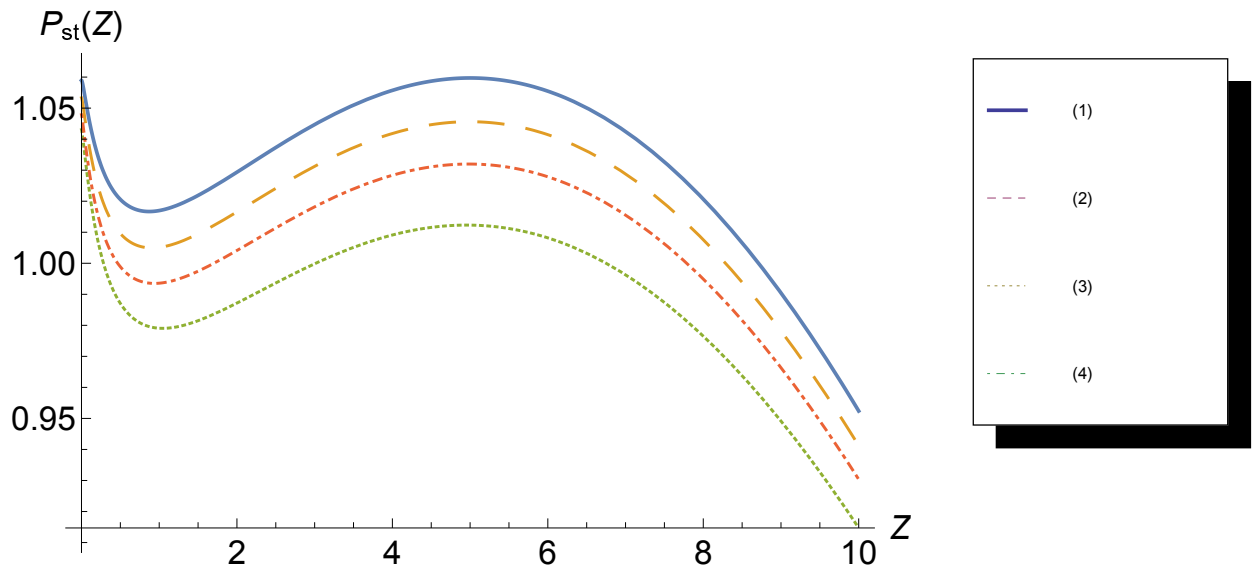}} &
 \end{tabular}
\caption{Probability distributions for model III with $g(z)=\frac{z+\gamma}{z+p_1}$ and with 
both additive and multiplicative noise (1) $D_1=0.03$, $D_2=1$ and $\lambda=0$  (2) $D_1=0.03$, $D_2=1$ and $\lambda=0.05$ 
(3) $D_1=0.07$ $D_2=1$ and $\lambda=0.05$  and (4) $D_1=0.03$ $D_2=1$ and $\lambda=0.1$. The other parameter values for all the curves are $p_0=0.28$, 
$\gamma=0.01$, $p_1=0.2$ $r_{bm}=0.22$ and $\gamma_z=0.01$. }
\label{fig:probdist_compare_transcrip_codegrade}
\end{figure}

 Figure (\ref{fig:probdist_compare_transcrip_codegrade}) shows how, for a given Z-sRNA synthesis  rate,  the noise strength and noise correlation influence the probability 
 distributions for model III with $g(z)=\frac{z+\gamma}{z+p_1}$.

\begin{figure}[ht!]
\begin{tabular}{cc}
\subfloat[$p_0=0.26$]{\includegraphics[width = 2.5in]{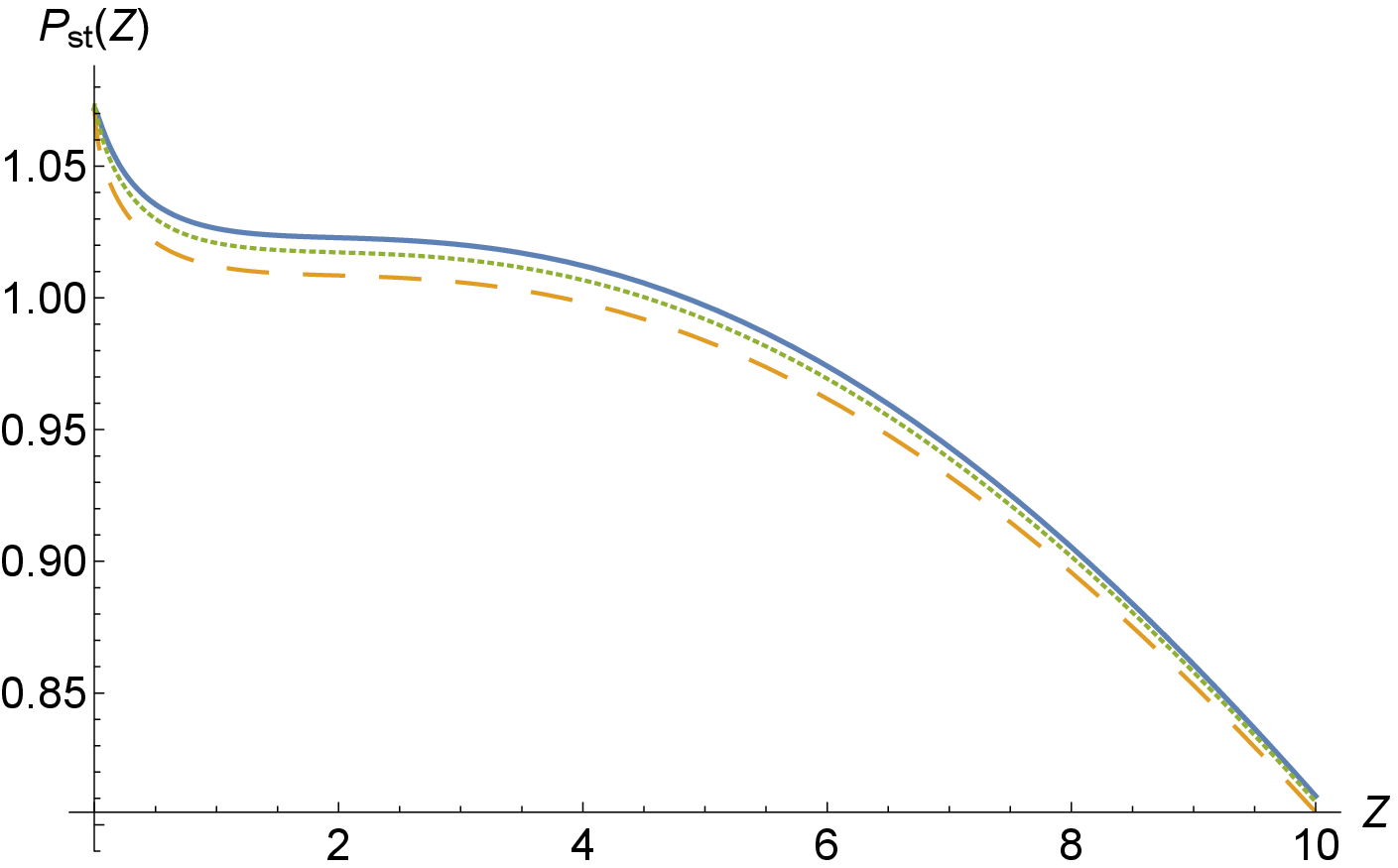}} &
\subfloat[$p_0=0.28$]{\includegraphics[width = 2.5in]{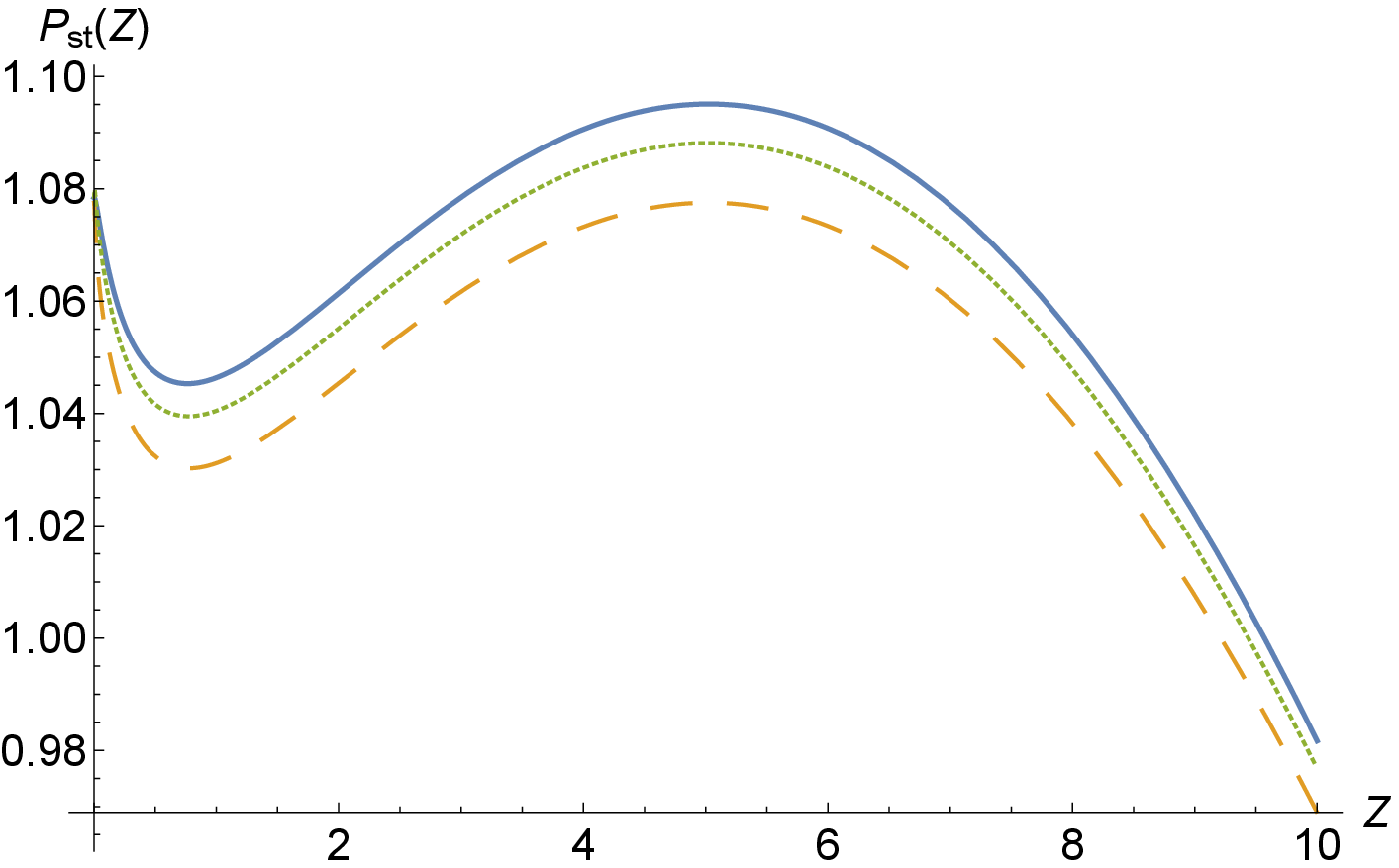}} 
\end{tabular}\\
\begin{tabular}{c}
\subfloat[$p_0=0.32$ ]{\includegraphics[width = 6.5in]{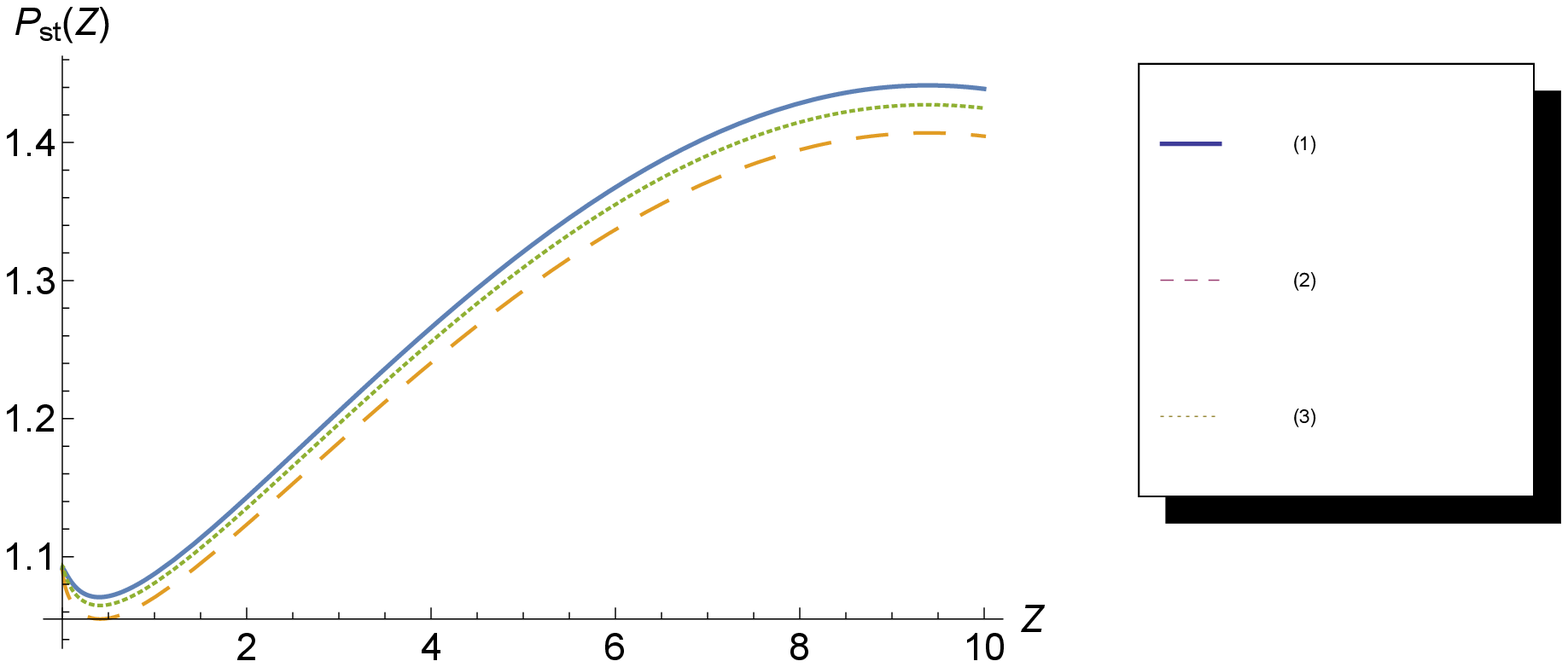}}
 \end{tabular}
\caption{Probability distributions for model III with  $g(z)=-\frac{z}{z+\gamma}$  and with (1) additive noise ($D_2=1$, $D_1=0$ and $\lambda=0$) 
(2) additive and multiplicative noise without correlation with ($D_2=1$,  $D_1=0.03$ and $\lambda=0$)
(3) additive and multiplicative noise with correlations with ($D_2=1$, $D_1=0.03$  and $\lambda=0.05$)
are plotted. These  plots are obtained with  $g(z)=-\frac{z}{z+\gamma}$.   
Other parameter values are $\gamma=0.01$, $p_1=0.2$, $r_{bm}=0.22$ and $\gamma_z=0.01$. }
\label{fig:alldist_translation_codegrade}
\end{figure}

Figure (\ref{fig:alldist_translation_codegrade})  shows  the 
probability distributions  for model III with  $g(z)=-\frac{z}{z+\gamma}$ for different synthesis rate of Z-sRNA. 

\pagebreak

\end{document}